\documentclass[aip, amsmath,amssymb,reprint]{revtex4-1}

\usepackage{graphicx}
\usepackage{dcolumn}
\usepackage{bm}
\usepackage[utf8]{inputenc}
\usepackage[T1]{fontenc}
\usepackage{mathptmx}
\usepackage{verbatim}
\usepackage{fdsymbol}
\usepackage{graphicx}
\usepackage{sidecap}
\usepackage{placeins}
\usepackage{xcolor}
\usepackage{amsfonts,amsthm,amsmath,amssymb,amscd,graphicx}
\usepackage{mathtools}
\usepackage{hyperref}

\makeatletter
\DeclareTextCommand{\textprime}{\encodingdefault}{%
  \mbox{$\m@th'\kern-\scriptspace$}%
}
\makeatother

\begin{document}

\preprint{AIP/123-QED}

\title[Homoclinic chaos in the R\"ossler model]{Homoclinic chaos in the R\"ossler model}

\author{Semyon Malykh}
\affiliation{National Research University Higher School of Economics, \\
             25/12 Bolshaya Pecherskaya Ulitsa, 603155 Nizhny Novgorod, Russia.}
\author{Yuliya Bakhanova}
\affiliation{National Research University Higher School of Economics, \\
             25/12 Bolshaya Pecherskaya Ulitsa, 603155 Nizhny Novgorod, Russia.}
\author{Alexey Kazakov}
\email{kazakovdz@yandex.ru}
\affiliation{National Research University Higher School of Economics, \\
             25/12 Bolshaya Pecherskaya Ulitsa, 603155 Nizhny Novgorod, Russia.}
\author{Krishna Pusuluri}
\affiliation{Neuroscience Institute, \\Georgia State University,
100 Piedmont Ave SE Atlanta, GA 30303, USA.}
\author{Andrey Shilnikov}
\affiliation{Neuroscience Institute and Department of Mathematics \& Statistics,\\ Georgia State University,
100 Piedmont Ave SE Atlanta, GA 30303, USA.}

\date{\today}

\begin{abstract}
We study the origin of homoclinic chaos in the classical 3D model proposed by O.~ R\"ossler in 1976. Of our particular interest are the convoluted bifurcations of the Shilnikov saddle-foci and how their synergy determines the global unfolding of the model, along with transformations of its chaotic attractors. We apply two computational methods proposed, 1D return maps and a symbolic approach specifically tailored to this model, to scrutinize homoclinic bifurcations, as well as to detect the regions of structurally stable and chaotic dynamics in the parameter space of the R\"ossler model.
\end{abstract}
\maketitle

\textbf{This paper is dedicated to Otto R\"ossler on the occasion of his 80th anniversary. He, being one of the pioneers in the chaosland, proposed a number of simple models with chaotic~\cite{Rossler1976,Rossler1979} and hyper-chaotic~\cite{Rossler1979hyp} dynamics that became classics in the field of applied dynamical systems. The goal of our paper is to examine and articulate the pivotal role and interplay of two Shilnikov saddle-foci \cite{Shilnikov1965} in the famous 3D R\"ossler model as they shape the topology of the chaotic attractors such as spiral, screw-type without and with funnels, and homoclinic, as well as determine their metamorphoses, existence domains and boundaries. Using the symbolic approach we biparametrically sweep its parameter space to identify and describe  periodicity/stability islands within chaoticity, as well as to examine in detail a tangled homoclinic unfolding that invisibly bounds the observable dynamics. The innovative use of 1D return maps generated by solutions of the model lets us quantify the complexity of chaotic attractors and provides a universal framework for the description of a rich multiplicity  of homoclinic phenomena that the classical R\"ossler model is notorious for.}


\section{Introduction}

The R\"ossler model~\cite{Rossler1976,Rossler1979} is a classic example of multi-faced deterministic chaos occurring in many low- and high-order systems. Its best known feature is the onset of chaotic dynamics due to a Shilnikov saddle-focus, that begins off with a period-doubling bifurcation cascade. The goal of this paper is to show how homoclinic bifurcations of two saddle-foci determine the organization and structure of chaotic attractors in the R\"ossler model.  We consider its following representation:
\begin{equation}
\dot x = -y - z, \quad \dot y = x + ay, \quad  \dot z = bx +z (x-c),
\label{eq:Rossler}
\end{equation}
 with $x,y$, $z$ being the phase variables, and $a,\,c>0$ being bifurcation parameters; we keep  $b=0.3$ fixed throughout this study. The convenience of the representation~(\ref{eq:Rossler}) is that  one equilibrium (EQ) state, called $O_1$, of the model is always located at the origin $(0,0,0)$, while the coordinates of the  second one $O_2$ are positive and given by $(c-ab, b - c/a, -(b - c/a))$. One can notice that when $c=ab$,  $O_2$ merges with $O_1$ and passes through it as a result of a transcritical saddle-node bifurcation.  This is not the case in the other form of the R\"ossler model \cite{Rossler1979}:
 \begin{equation}
\dot x = -y - z, \quad \dot y = x + ay, \quad  \dot z = b +z (x-c),
\label{eq:Rossler1}
\end{equation}
 where the location of neither equilibrium is fixed, and they emerge in the phase space through a generic saddle-node bifurcation.

  To become a saddle-focus, $O_1$ loses stability through a super-critical Andronov-Hopf (AH) bifurcation. Note that the transcritical saddle-node bifurcation makes the other EQ $O_2$ a saddle-focus from the very beginning. It also undergoes an Andronov-Hopf bifurcation, though sub-critical after a saddle periodic orbit collapses into  $O_2$ to convert it into a repeller. We will proceed with this discussion  in the context of homoclinic bifurcations below.

 \begin{figure*}[t]
\includegraphics[width=0.75\linewidth]{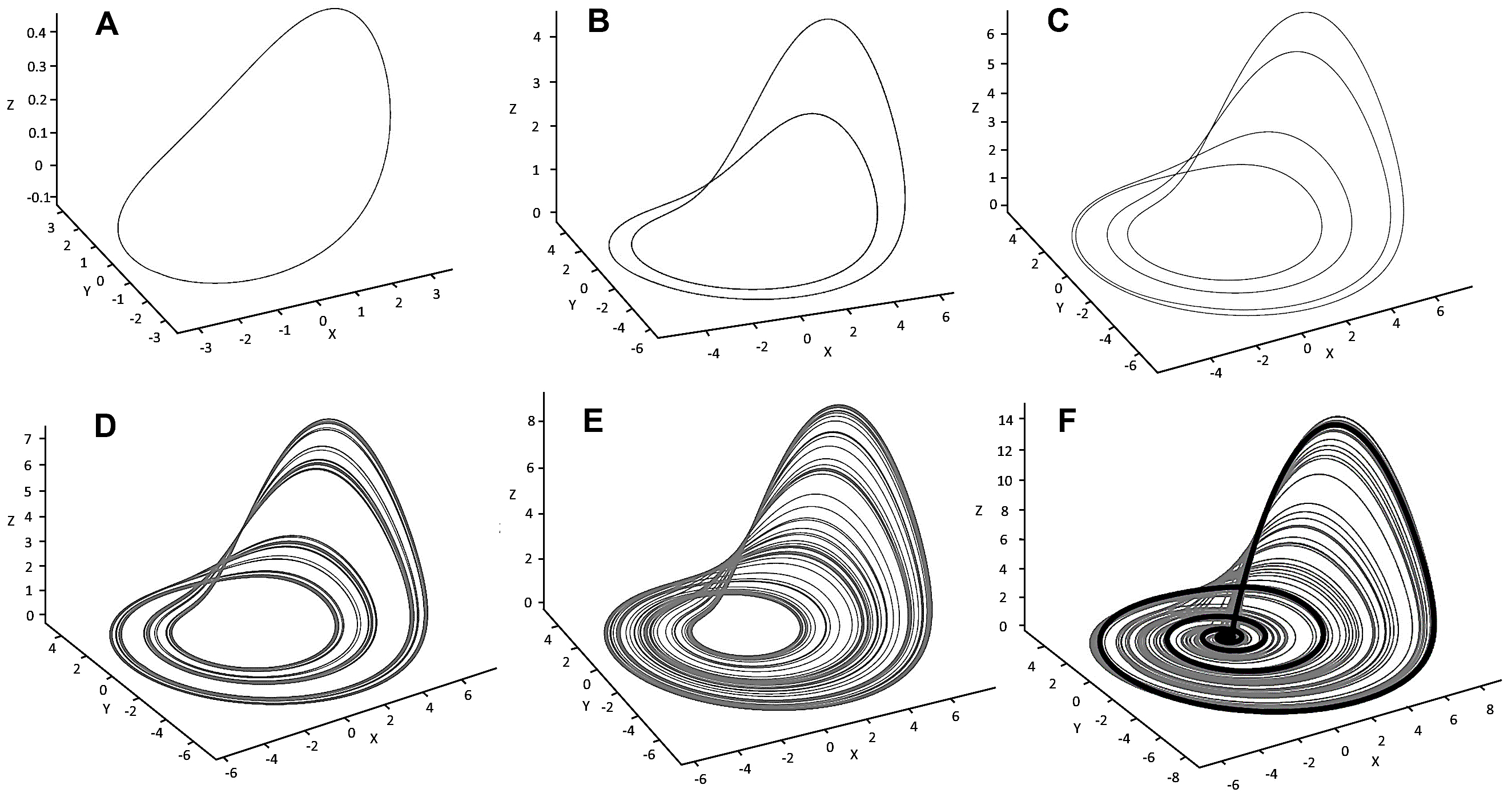}
\caption{Period-doubling bifurcations en route to spiral chaos in the 3D phase space of the Rossler model on a pathway at $c=4.9$. (A) Stable periodic orbit (PO) at $a = 0.08$ is followed by a period-2 orbit at $a = 0.25$ in (B) and next by a stable period-4 obit at $a = 0.275$ in (C). Period-doubling chaotic attractors at $a = 0.28$  and $0.3$ shown, resp., in panels D and E morph into a spiral attractor with the Shilnikov homoclinic saddle-focus $O_1$ embedded at $a = 0.35$ (F), after the shrinking hole around it fully collapses.}
\label{fig:pd}
\end{figure*}

 The AH bifurcation transforms the stable equilibrium $O_1$ into a saddle-focus of type the (1,2), i.e., with 1D stable and 2D unstable manifolds due to a single negative eigenvalue   $\lambda_1 < 0$, and a pair of complex-conjugated eigenvalues $\lambda_{2,3} = \alpha  \pm i \omega$  with a positive real part ($\alpha>0$), see insets in Fig.~\ref{fig:Rossler_BifDiagr}B.  Vice versa, the saddle-focus $O_2$ is of the (2,1) type.

 The vast majority of the papers on the onset of properties of chaos occurring in the R\"ossler model are focused on bifurcations related to the primary EQ $O_1$, while leaving the role of the secondary EQ $O_2$ in the shadow. We will reveal its contribution, through homoclinic bifurcations, to the overall global chaotic dynamics and its transformations in \eqref{eq:Rossler}.

 Let us point out that in the parameter space of interesting dynamics for the model, the $c$-parameter is greater on the order of magnitude than the other parameters, $a$ and $b$ in the $(0,1)$-range, see details in \cite{rossler06scholarpedia,rosslerWiki}.

 The discrepancy in the parameter values implies that the Eqs.~\eqref{eq:Rossler} have two time scales with two slow $(x,y)$- and one fast $z$-phase variables. Moreover the divergence of the vector field generated by Eqs.~\eqref{eq:Rossler} is estimated by $a+x-c$ or $a-c<0$ around the origin $O_1$. This makes the R\"ossler model strongly dissipative around the origin of the 3D phase space. However, that is not the case near the other equilibrium state $O_2$ where the divergence of the vector field is small and positive: $a(1-b) \sim o(2)$. This observation partially explains a slow convergence to $O_2$ along its 2D stable manifold as observed from  Figs.~\ref{fig:spiral_screw_funnel}B$_1$-C$_1$.

 \subsection*{Period-doubling cascade to spiral chaos} \label{sec:scenario}

Let us recap without excessive details what is well-known: the route to spiral chaos in the R\"ossler model through a cascade of period-doubling bifurcations.  It is documented in Fig.~\ref{fig:pd} with fixed $c = 4.9$ as the $a$-parameter is increased. The supercritical AH-bifurcation makes  $O_1$ the saddle-focus and gives rise to the emergence of a stable periodic orbit (PO), see Fig.~\ref{fig:pd}A. With a further increase in $a$, this PO loses the stability inherited by an orbit of period-2 and next by an orbit of period-4 (Figs. \ref{fig:pd}B-C) through two initial
period-doubling bifurcations. Further steps lead to the onset of a longer PO and a Feigenbaum-type
strange attractor discovered by O.~R\"ossler in the model named after him \cite{Rossler1976}. Its interaction with a transverse cross-section  in 3D would look like a  H\'enon attractor \cite{Henon76} with a recognizable parabola shape as one depicted in Fig.~\ref{fig:mapScheme}. Observe the shrinking hole around $O_1$  as depicted in  Figs.~\ref{fig:pd}D-E indicative that the saddle-focus remains yet isolated. Eventually, at $a \approx 0.35$ it collapses and leads to the formation of the primary homoclinic loop of $O_1$, see Fig.~\ref{fig:pd}F. It is easy to argue that the Shilnikov bifurcation \cite{Shilnikov1965}  causes a homoclinic explosion (hyperbolic subset)  with countably many saddle POs inside the so-called {\em Shilnikov whirlpool} \cite{OSh86e} formed nearby $O_1$ in the phase space. It follows from the above arguments concerning the divergence that $\sum_{i=1}^3 \lambda_i=c-a>0$ for the Eqs.~(1) in backward time,  which implies that the Shilnikov condition: the saddle value $\lambda_1-\alpha>0$, or the saddle index $\rho=\lambda_1/\alpha >1$, is fulfilled. Saddle orbits remain saddle in forward time too. This explains the nature of chaotic dynamics near the Shilnikov saddle-focus. A historic remark: the paper  \cite{ArnCoulTres1982} by Arneodo, Collet and Tresser  was the first numerical evidence of strange attractors associated with the Shilnikov theorem in the R\"ossler model~\eqref{eq:Rossler}.

 Recall, however, that a 3D dissipative system with the Shilnikov saddle-focus cannot produce a purely chaotic attractor but a quasi-attractor \cite{AfrShil1983,GST97}, due to homoclinic tangencies that give rise to stable POs unpredictably emerging in its phase-space through saddle-node bifurcations and followed by period-doubling ones. To detect such stable POs and to determine the corresponding stability windows in the biparametric sweeps, we propose and develop a symbolic approach combined with a phase space partition. With this approach, one can generate long binary sequences whose Lempel-Ziv complexity \cite{lempel1976complexity} indicates whether they correspond to regular or chaotic dynamics in the model. On the contrary, short binary sequences let us detect homoclinic bifurcations of saddle-foci in the phase and parameter spaces.

  \begin{figure*}[ht!]
\center{
\includegraphics[width=.9\linewidth]{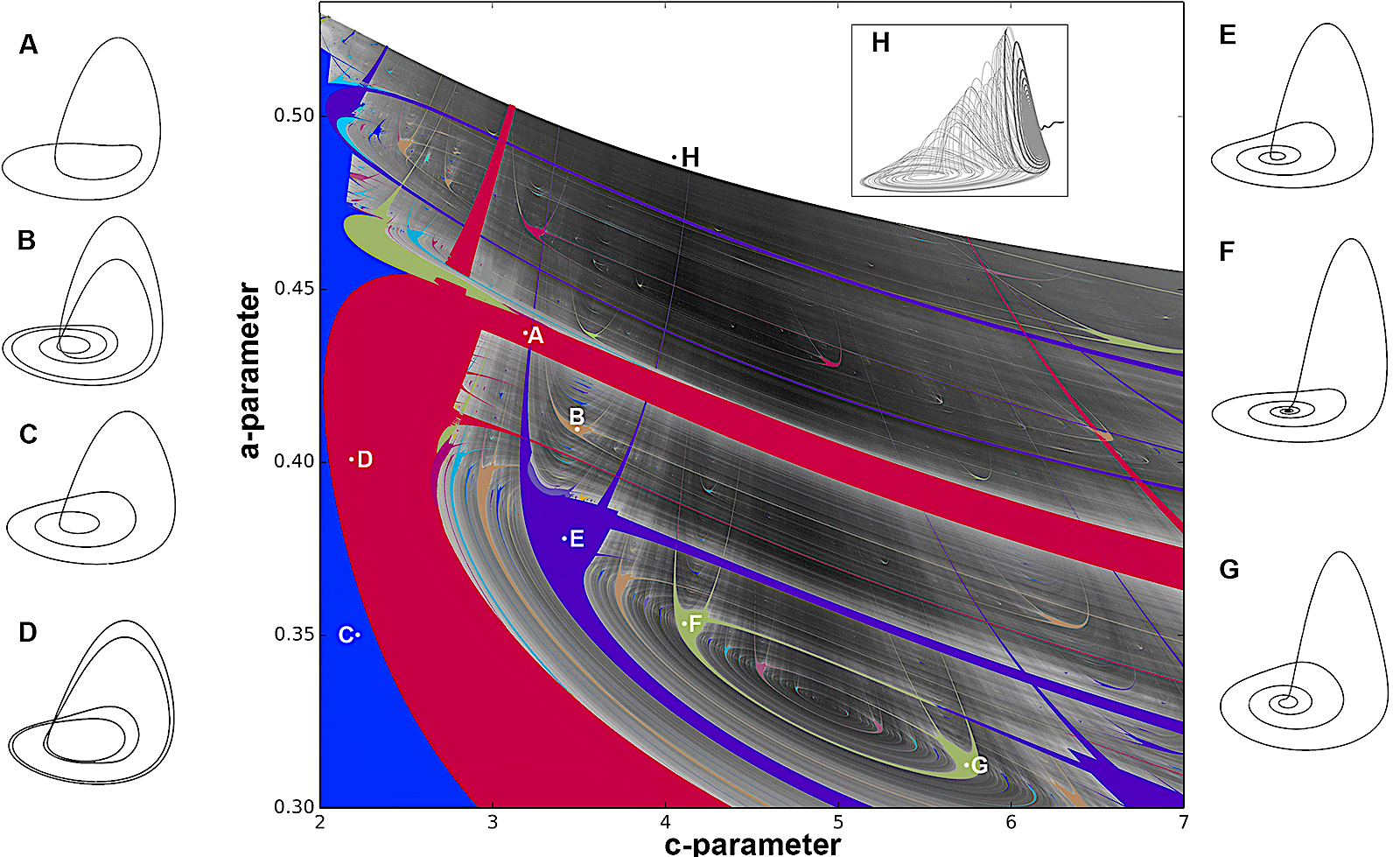}}
\caption{The ($c,\,a$)-parameter [$5000 \times 5000$]-resolution sweep of the R\"ossler model with the aid of DCP and LZ-complexity approach utilizing 1000-symbols long sequences, with first 1000 transients omitted. Shown in solid colors are multiple stability islands, including the so-called {\em shrimp}-shaped regions, hosting stable periodic orbits inside (some sampled in panels
A-G), while the regions in the grey-ish color are associated with chaotic dynamics  (H) Solutions of \eqref{eq:Rossler} start escaping to infinity above some demarcation line (discussed in detail below).
}
\label{fig:Rossler_LZdiagr}
\end{figure*}

 The paper is organized as follows: firstly, we scan the parameter plane to detect the regions of regular and chaotic dynamics in the R\"ossler model. That is done by partitioning the 3D phase space and introducing a symbolic description for long-term solutions of Eqs.~(1). We next analyze the collected binary sequences to determine whether they are periodic or aperiodic and what the Lempel-Ziv complexity of chaotic sequences is. This fast and effective approach is an alternative to a less speedy method of Lyapunov diagrams, see details in Sec.~\ref{sec:LZmethod}. The following Section~\ref{sec:1DMaps} is focused on the proposed  algorithms for computationally devising 2D and 1D return maps  to examine the structure of the chaotic attractors in the model, and what makes them morph from
 spiral to multi-funnel shapes and how to quantify the transition in the parameter space.
We argue that the number of critical points in such 1D  map can be used to reasonably categorize homoclinic chaos associated with the saddle-foci as spiral, screw, and multi-funnel attractors, see details in Sec.~\ref{sec:1DMaps}. Next we discuss an efficient approach to search for a small family of homoclinic bifurcations of the primary saddle-focus $O_1$ in biparametric sweeps. Having found these bifurcation curves, we employ the developed machinery of the 1D maps to give insights into a fine structure of  special cases -- hubs where homoclinic curves form the edges of the chaotic attractors in the phase space. In addition, Section~\ref{sec:1DMaps} highlights how 1D maps can predict the homoclinic bifurcations before they occur in the R\"ossler model. Section \ref{sec:ShrimpsAndHubs} is focused on the organization of a typical hub and how several periodicity hubs shape the stability islands in the parameter space. Section~\ref{sec:symbols} discusses how the solution of the R\"ossler model becomes unbounded and what is the pivotal role of the secondary saddle-focus in shielding them from escaping to infinity. We reveal that the crisis is due to invisible homoclinic bifurcations of the saddle-focus $O_2$ and how this is related such that the R\"ossler model becomes no longer dissipative but expanding. Using the short symbolic dynamics we find the first basic homoclinic bifurcations of $O_2$ and demonstrate how they arrange and demarcate the existence region of deterministic chaos in the given model. In Section \ref{sec:h1Hyp}, we will speculate about a nested organization of homoclinic bifurcation curves in the interior of the primary U-shaped one.

\begin{figure*}[hbt!]
\center{\includegraphics[width=.6\linewidth]{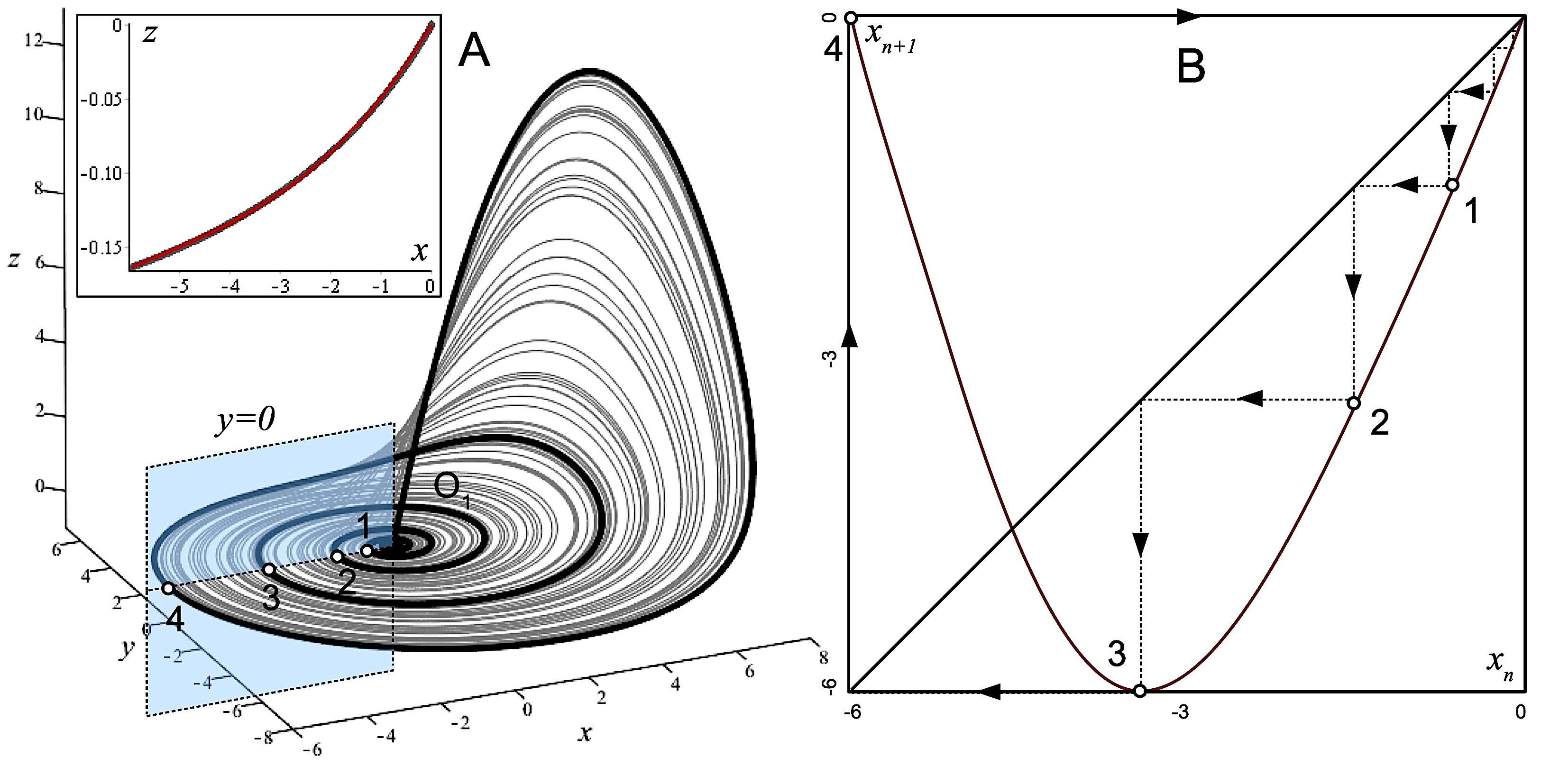} }
\caption{(A) Spiral attractor ($c=4.93$, $a=0.33$), superimposed with the primary homoclinic orbit of the saddle-focus $O_1$, and its intersection points with the 2D cross-section (blue plane) given by $y=0$ and $x \leq 0$.  (B) Computationally interpolated 1D return map $F:  x_n \to x_{n+1}$ of an $x$-interval spanning from $O_1$ through the edge of the chaotic attractor. Forward and backward iterates, $\{1, 2, 3, 4\}$, of the critical point on the $x_n$-axis, approach/converge, resp., to the repelling fixed point (FP) at the origin $O$, corresponding to the homoclinic orbit of $O_1$ in (A). }
\label{fig:mapScheme}
\end{figure*}

 \section{Biparametric sweep with  LZ complexity and deterministic chaos prospector} \label{sec:LZmethod}

In the given section, we discuss the symbolic approach combined with the partition of the phase space to detect  the regions of chaos and stability islands in the parameter space of the R\"ossler model. This is the first step prior to applying more dedicated tools for examining a variety of homoclinic bifurcations. We previously developed a symbolic toolkit, code-named deterministic chaos prospector (DCP), running on graphics processing units (GPUs) to perform in-depth, high-resolution sweeps of control parameters to disclose the fine organization of characteristic homoclinic and heteroclinic bifurcations and structures that have been  universally observed in various Lorenz-like systems, see  \cite{xing2015fractal, Xing2014, Barrio2013,pusuluri2018homoclinic, pusuluri2017unraveling} and the reference therein. In addition to this approach capitalizing on sensitive dependence of chaos on parameter variations, the structural stability of regular dynamics can also be utilized to fast and accurately detect regions of simple and chaotic dynamics in a parameter space of the system in question \cite{pusulurihomoclinicCNSNS}. The $Z_2$-symmetry is exploited to generate periodic or aperiodic binary sequences that can be associated, respectively, with simple or chaotic flip-flop patterns of solutions of Lorenz-like systems. The application of the symbolic techniques for  studies of neural dynamics in cell models and small neural networks is demonstrated in \cite{pusuluri2019symbolic, Pusuluri2020Chapter, pusuluri2020dissertation}.

 Our practical approach tailored specifically for homoclinic chaos in the R\"ossler model uses
particular events when the $z$-variable reaches consequently its maximal values  on the attractor, see Fig.~1F. The binary sequence $(k_n)$ representing a trajectory is computed as follows:
\begin{equation}\label{kneadingsMain}
k_n = \begin{dcases*}
1, & when $z_{max}$ > $z_{\rm th}$, \\
0, & when $z_{max}$ $\leq$ $z_{\rm th}$.
\end{dcases*}, \quad  z_{\rm th} = 0.12 (c - ab)/a.
\end{equation}
Here, the $z$-threshold is set relative to the location of the saddle-focus $O_2$. The choice of partition is motivated by its simplicity and may possibly be different, yet sufficient for our purpose. We note that various partitioning approaches may be employed for the same end result - effective detection of stability windows corresponding to regular and structurally stable dynamics in the parameter space of the R\"ossler model.

With this simple algorithm we map the maxima values of the $z$-variable to the binary symbols $0$ and $1$, based on some threshold value $z_{th}$. Biparametric sweeps such as those shown Figs.~\ref{fig:Rossler_LZdiagr}, \ref{fig:Rossler_BifDiagr}A, \ref{fig:LZ_hubs}A are obtained by computing long trajectories starting from identical initial conditions (near the origin), and by skipping some initial transients,  as two control parameters, $c$ and $a$, are varied across a $5000 \times 5000$-size grid.

We use the fourth order Runge-Kutta method with fixed step-size (GPU requirement) for numerical integration. The computation of trajectories across different parameter values is parallelized using GPUs. Data visualization is done using Python. Long term system dynamics are analyzed by omitting the first $1000$ symbols as transients, and processing the following $1000$ symbols, to detect periodicity within the sequence. An overbar is used to represent periodic sequences such as $(010101...)$ by simply $(\overline{01})$. Aperiodic strings representing chaotic trajectories, are processed using the Lempel-Ziv (LZ) compression algorithm to measure their complexity \cite{lempel1976complexity}. As a string is continuously scanned, new words are added to the LZ vocabulary. The size of the vocabulary towards the end of the string, normalized with its length  is used as the complexity measure, which is low for periodic sequences and high for chaotic ones. Chaotic dynamics are associated with the region painted in biparametric sweeps in the shades of gray, so greater LZ-complexity means darker gray, thus indicating higher instability. For trajectories mapped to periodic strings, the shift symmetry of the sequences must be considered.  For example, depending on the length of the initial transient omitted, the same periodic orbit could be represented by either $\{\overline{01}\}$) or $\{\overline{10}\}$. We normalize such shift symmetric periodic sequences to the smallest binary valued circular permutation. Thus, both $\{\overline{01}\}$ and $\{\overline{10}\}$ are normalized to $\{\overline{01}\}$, while the periodic sequences $\{\overline{011}\}$, $\{\overline{110}\}$, $\{\overline{101}\}$ are normalized to $\{\overline{011}\}$. Let us re-iterate that we deliberately tailor this approach only to examine and detect chaos and stability islands due to saddle-node bifurcations in the parameter sweep. As such, period-doubling bifurcations are beyond the scope of  our consideration, even though the symbolic approach can be further enhanced to detect such bifurcations as well.

 \begin{figure*}[hbt!]
\center{\includegraphics[width=.8\linewidth]{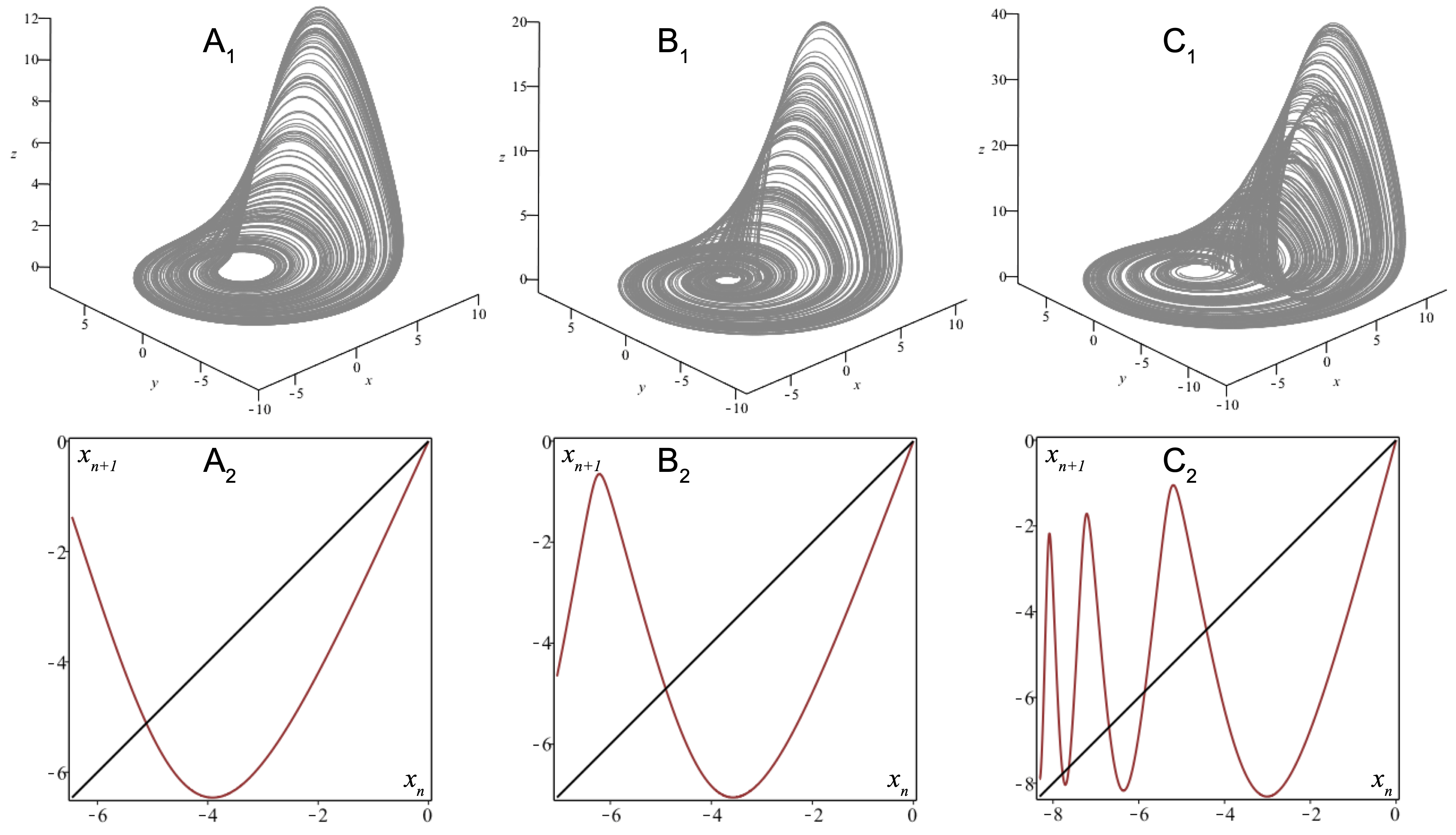}}
\caption{Changes in the topology of the chaotic attractor in the R\"ossler model as it shifts closer to the secondary saddle-focus $O_2$ along the pathway at $c=5.5$. (A$_1$) the spiral attractor at $a=0.3$  and (A$_2$) the corresponding 1D unimodal map; (B$_1$) a screw-like attractor at $a=0.355$, and (B$_2$) the corresponding 1D map with two critical points; and (C$_1$) multi-funnel attractor at $a=0.455$ and (C$_2$) the associated 1D map with multiple critical points. As the saddle-focus $O_1$ is isolated from the non-homoclinic chaotic attractor in each phase projection, all maxima in the corresponding maps are located below the repelling FP at $O$ representing $O_1$.}
\label{fig:spiral_screw_funnel}
\end{figure*}

In order to identify various stability windows in the parametric plane, we define a formal convergent power series $P$ for the normalized periodic sequence $\{k_i{\}}_{i=p}^q$ given by:
\begin{equation}\label{kneadingsSum}
P = \sum_{i=p}^{q} \frac{k_i}{2^{q+1-i}}.
\end{equation}
By construction, the $P$-value ranges between 0 and 1. The boundary values are set by the periodic sequences $\{\overline{0}\}$) and $\{\overline{1}\}$, respectively, for infinitely long sequences. Here, we employ $p=1000$ and $q=2000$. Parameter values that produce identical periodic orbits in their trajectories, result in identical sequences \(\{k_i\}\) (after normalization of shift symmetry), and thus, have the same $P$ values, which are then projected on to the biparametric sweep using a colormap. This colormap takes $P$ values into $2^{8}$ discrete bins of RGB-color values assigned from 0 through 1 for each of red, green and blue in decreasing, random and increasing order, respectively. The results of the sweep are demonstrated in Figs.~\ref{fig:Rossler_LZdiagr}, \ref{fig:Rossler_BifDiagr}A and \ref{fig:JointSweep}A. Stability windows with distinct periodic orbits can be painted with different or same solid colors in the sweeps, while chaotic regions of structural instability are shown in gray; moreover we re-emphasize the darker grey pixels are associated with more developed chaos in the R\"ossler model. To conclude this section, let us point out at the mysterious demarcation line separating the region (here in white) where the solutions of Eqs.~(\ref{eq:Rossler}) start escaping to infinity. The rest of the paper is basically dedicated to disclosing what this boundary is and how it  influences the evolution of chaotic dynamics with underlying homoclinic bifurcations of both saddle-foci.
 We will proceed with the discussion of these results in detail below.


\section{From spiral to screw-type chaos with funnels}\label{sec:spiral_fun}

The goal of this section is to demonstrate how chaotic attractors initially developed through period doubling cascades evolve further, specifically as they approach the demarcation line in Fig.~\ref{fig:Rossler_LZdiagr}. We will demonstrate the role and applicability of L.P. Shilnikov saddle-focus bifurcation and its unexpected developments \cite{Shilnikov1986} for the R\"ossler model.

It was well-reported and discussed that chaotic attractors in the R\"ossler model can
be of various shapes and complexity based on the number of nested sub-funnels,
see Fig.~\ref{fig:spiral_screw_funnel}, that increases as the demarcation line
is approached from below. Alternatively, the transition after which the solutions  become unbounded is often
associated with the crisis occurring in the model on the demarcation line \cite{barrio2009}. It is also well-known that the number of funnels can be effectively exposed with the use of 1D return map due to the strong contraction, as was done for the first time for the given system in \cite{GaspardKapralNicolis84}. 1D maps happened to be instrumental in determining the topology of the attractors in question \cite{LetellierDutertreMaheu95} and the complexity of saddle periodic orbits embedded into an attractor. Depending on the number of branches in the corresponding maps, chaotic attractors can be categorized as spiral, screw, and multi-funnel attractors \cite{Rossler1979, FraseKapral82}.


\begin{figure*}[htb!]
\center{\includegraphics[width=.9\linewidth]{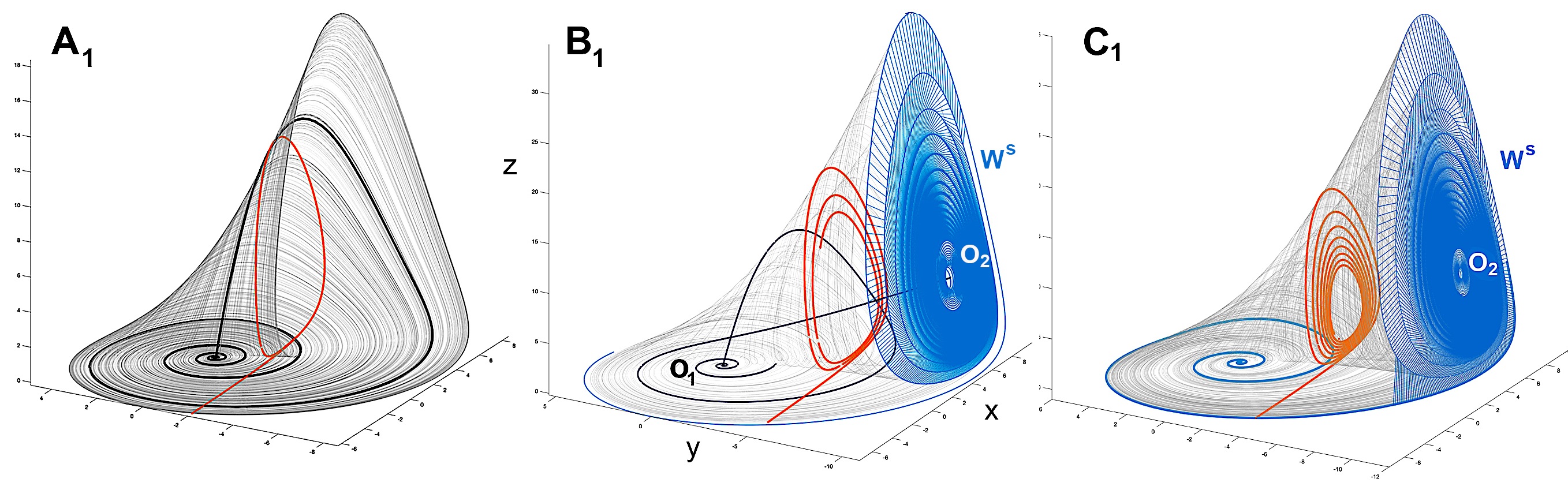}}
\caption{Chaotic attractors (gray color) morphing from (A$_1$)  the spiral one with a homoclinic orbit of the saddle-focus $O_1$ at $(c,\,a)\simeq (4.822,\,0.38)$  to the attractor with emerging funnels at $(4.615,0.468)$ with a unidirectional heteroclinic connection from $O_2$ to $O_1$ in (B$_1$) , and through the multi-funnel attractor at $(4.652,\,0.4775)$ in (C$_1$).
The spirals in the orange color reveal the ways the attractors hit some transverse cross-sections ($y=-2,$ -4) in the phase space. Blue lines in panels B$_1$ and C$_1$ represent the location of the 2D stable manifold $W^s$ of the saddle-focus $O_2$ that makes the chaotic attractor wraps  around the left unstable separatrix $W^{u+}$ of $O_2$ the more, the closer it shifts to the secondary saddle-focus, whose manifold $W^s$ shields the solutions of Eqs.~(1) from escaping to infinity, see Fig.~\ref{fig:Rossler_LZdiagr}H.}
\label{fig:Attractor_StableManifold}
\end{figure*}

\subsection*{Computational derivation of 1D maps}\label{sec:1DMaps}

Let us itemize our construction procedure for 1D return maps tailored for the R\"ossler attractors. First,  the half-plane $\{y=0,\, x\leq0\}$ is chosen as a cross-section transverse to spiraling trajectories of  \eqref{eq:Rossler} near $O_1$ to define a 2D Poincar\'e map $(x_{n+1}, z_{n+1}) = F(x_n,z_n)$, as illustrated in Fig.~\ref{fig:mapScheme}A. Due to the strong contraction near the saddle-focus $O_1$, the intersection points of the chaotic attractor with the cross-section appear to lie on a 1D densely populated curve, as depicted by the inset of Fig.~\ref{fig:mapScheme}A. Next, we parameterize this curve by the Lagrange polynomial $z(x)$ using four points: the first point $(0,0)$ corresponds to the saddle-focus $O_1$, the second point $(x_{min},z(x_{min}))$ is a point with the minimal $x$-coordinate belonging to the attractor, and two middle points on it. Finally, we evaluate the corresponding 1D map $x_{n+1} = F(x_n, z(x_n))$ using 5000 $x$-samples on the interval $[x_{min, 0}]$. The resulting one-dimensional map for the chaotic attractor embedded with the primary homoclinic orbit to the saddle-focus $O_1$ is presented in Fig.~\ref{fig:mapScheme}B. By construction, such a 1D map has a few pivotal features: (a) it inherits  a unimodal shape (with a single point) from a period-doubling cascade; (b) $0=F(0)$ is a repelling fixed point (FP) representing the saddle-focus $O_1$; (c) the far-right critical point always touches the $x_n$-axis; (d) in the homoclinic case, the forward iterates of the critical point  terminate at the repelling FP, while  its backward iterates converge to it along the right increasing branch of the map (corresponding to the homoclinic loop in Fig.~\ref{fig:mapScheme}A); (e) otherwise, the forward iterates of the critical point cannot reach the FP as the left branch(es) and all local maxima are lowered below, prior to or after the homoclinic bifurcation. This can be interpreted that the saddle-focus $O_1$ becomes (temporarily) isolated from the chaotic attractor in both super- and sub-critical homoclinic cases, like the one depicted in Fig.~\ref{fig:pd}E.

Having developed the proposed 1D map methodology, we can better elaborate on the classification of attractors by the complexity of the funnel proposed in \cite{Rossler1979, FraseKapral82, LetellierDutertreMaheu95}. Namely, whenever the corresponding 1D map has only two branches, the attractor is called spiral, see Fig.~\ref{fig:spiral_screw_funnel}A. It can be homoclinic if the critical point is mapped onto the FP at $O$, like in Fig.~\ref{fig:mapScheme}B, or not (when the left branch is not as high as the right  one, see Fig.~\ref{fig:spiral_screw_funnel}A$_2$). The chaotic attractor is called of screw-type when the map gains an additional (third) branch, see Fig.~\ref{fig:spiral_screw_funnel}B$_2$, and multi-funnel if the  map has four or more branches, see Fig.~\ref{fig:spiral_screw_funnel}C$_2$. We re-emphasize that either attractors become homoclinic only if the images (forward iterates) of the far right critical point on the $x_{n+1}=0$-axis end up at the repelling FP at the origin.

In conclusion, we would like to say that the proposed 1D map approach not only helps to differentiate the chaotic attractors by the complexity of the funnels, but can also predict and accurately verify whether any homoclinic loop of the saddle-focus  $O_1$ occurs or not for the given parameter values.

\subsection*{Funnels stirred by saddle-focus $O_2$}\label{funnels}

Let us highlight an invisible role of the second saddle-focus $O_2$ implicitly regulating the formation of funnels in the chaotic attractors in the R\"ossler model. Recall that this saddle-focus is of the (2,1)-type. Its 2D manifold $W^{s}$, locally dividing the 3D phase, forms an umbrella shielding the domain of the chaotic attractor that prevents nearby solutions from escaping its attraction. Its 1D unstable  manifold $W^u$ includes $O_2$ itself and two outgoing separatrices, say $W^{u+}$ and $W^{u-}$, such that  $W^{u+}$ converges to an attractor (Fig.~\ref{fig:Attractor_StableManifold}B$_1$), while $W^{u-}$ tend to infinity (Fig.~\ref{fig:Rossler_LZdiagr}H).

Figures~\ref{fig:Attractor_StableManifold}B$_1$ and \ref{fig:Attractor_StableManifold}C$_1$ illustrate the role of $O_2$ in the funnel formation. Both depict the
chaotic attractor in the phase space as the demarcation line in Fig.~\ref{fig:Rossler_LZdiagr} is approached from below in the parameter space. The attractor is superimposed with a local portion of the stable manifold (shown in blue) of the saddle-focus $O_2$; blue spirals also reveal the slow rate of convergence to $O_2$. The (orange) spiraling line is composed of a large number of intersection points of the chaotic attractor with a transverse cross-section ($y=-2$ or $-4$). One can notice that the closer the attractor gets to the stable manifold of $O_2$, the more it becomes wrapped around $W^{u+}$.  Figure~\ref{fig:Attractor_StableManifold}B$_1$ also shows $W^{u+}$ unidirectionally connecting $O_2$ with $O_1$. This explains the origin of the funnels observed in the chaotic attractors in the R\"ossler model, which are reflected in adding new branches in the 1D return maps (Figs.~\ref{fig:spiral_screw_funnel}A$_2$-C$_2$).


\begin{figure*}[ht!]
\center{\includegraphics[width=.99\linewidth]{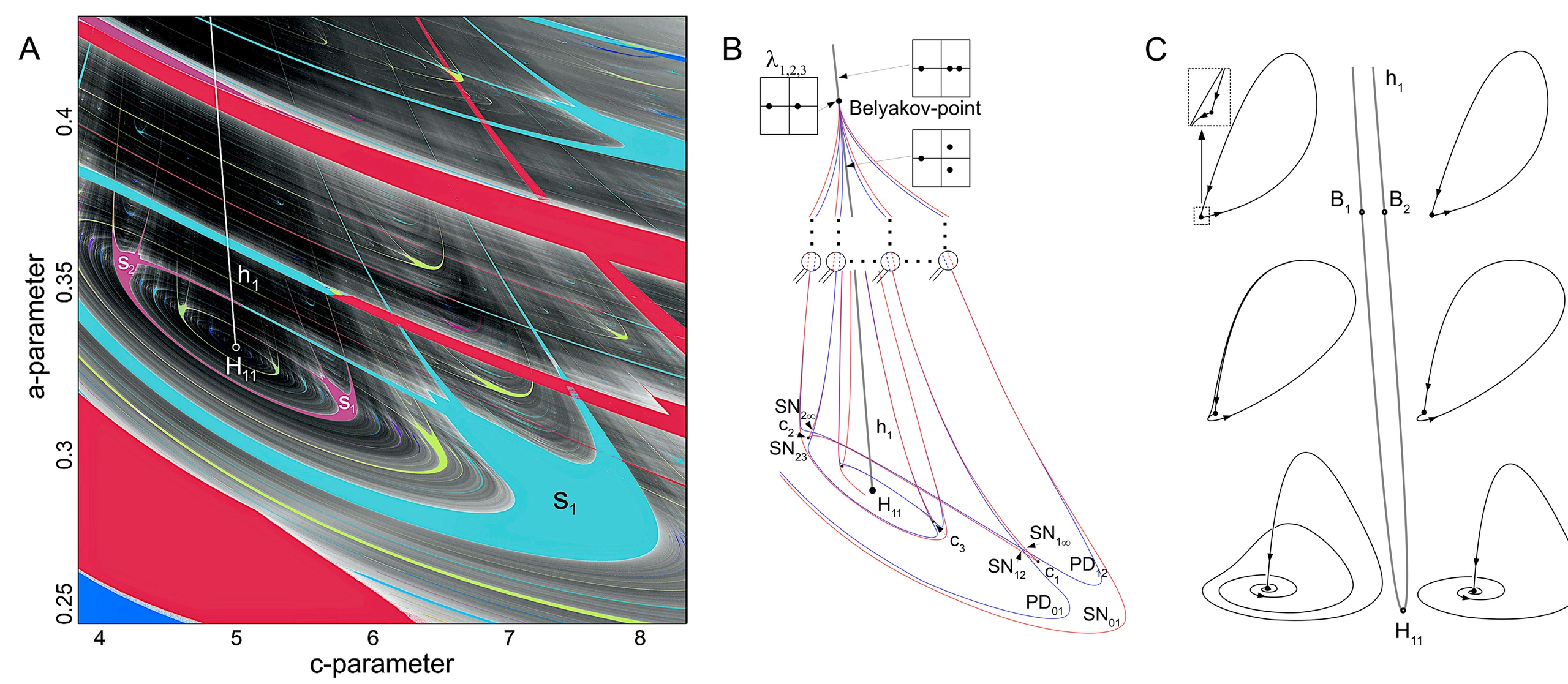} }
\caption{  (A) Fragment of the biparametric sweep. Solid, colored regions S$_1$, S$_2$, S$_3, \ldots$ are associated with stability islands corresponding to the attracting POs, emerging within chaos-land painted with the gray-ish colors. The line h$_1$ with the end-point H$_{11}$ corresponds to the primary homoclinic bifurcation of the saddle-focus $O_1$ (see Fig.~\ref{fig:pd}F).
(B)   Sketch of the self-similar bifurcation diagram near h$_1$.  Outer borderline of a shrimp-shaped region S$_i$ are due to saddle-node (SN) bifurcation curves SN$_{i-1,i}$ (in red) bridging with a successful one S$_{i-1}$ and originating from the Belyakov cod-2 point corresponding to a homoclinic saddle with double real eigenvalues $\lambda_1=\lambda_2<0$. Inner demarcation line of $S_i$ is due to period-doubling (PD) bifurcations (blue lines): the first curve PD$_{i-1,i}$ connecting successive S$_{i}$ and S$_{i-1}$ regions also starts off the Belyakov point, and so forth. Note the cusps on SN-curves such that SN$_{i,i+1}$ links S$_i$ and S$_{i+1}$ islands, while SN$_{i,\infty}$ terminates at the Belyakov point. (C) Evolutions of single and double homoclinic orbits to $O_1$  transitioning from a saddle to a saddle-focus  and back along the U-shaped bifurcation curve h$_1$,  with the tip H$_{11}$ being a turning point (simulations due to MatCont~\cite{matcont2008, marcont2012}).}
\label{fig:Rossler_BifDiagr}
\end{figure*}

\section{Homoclinics, hubs and shrimps}\label{sec:ShrimpsAndHubs}

Here we will discuss phenomena that underlie a multiplicity of periodicity windows (well seen in  Fig.~\ref{fig:Rossler_LZdiagr}). It is very well-known  \cite{Shilnikov1965, OSh86e, a2001methods, Shilnikov_heritage, Shilnikov_scolarpedia} that there are countably many saddle POs near the Shilnikov homoclinic saddle-focus in the phase space. The  stable and unstable manifolds of such saddle POs can as cross each other transversally as become tangent;  moreover, systems with such homoclinic tangencies fill in open regions of the parameter space. Specifically, homoclinic tangencies give rise to the emergency of stable POs in the 3D phase space through forthcoming saddle-node bifurcations \cite{gavrilov1972three, gavrilov1973three, GST97} see a few stable OPs sampled in Fig.~\ref{fig:Rossler_LZdiagr}A--\ref{fig:Rossler_LZdiagr}G. One can see from the sweep in \ref{fig:Rossler_LZdiagr} that the parameter space of the R\"ossler model, which is mostly populated with chaotic dynamics also embeds solid-colored regions called periodicity  windows or stability islands that are populated by attracting orbits. Strange attractors of such intermediate nature, where hyperbolic subsets may coexist with stable POs therein, were termed \textit{quasi-attractors} by V.S. Afraimovich and L.P. Shilnikov  \cite{AfrShil1983}, see also \cite{GST97}. Below, we will elaborate on the bifurcation unfolding of those self-similar stability windows, labelled by S$_i$ in Fig.~\ref{fig:Rossler_BifDiagr}A, that have the specific shrimp-shape.

The gray curve h$_1$ in Fig.~\ref{fig:Rossler_LZdiagr}A corresponds to the primary homoclinic bifurcation of the saddle-focus equilibrium $O_1$. The study \cite{GaspardKapralNicolis84} was the first where the bifurcation analysis of stability windows emerging near this curve was carried out in detail. Its unfolding is outlined in Fig.~\ref{fig:Rossler_LZdiagr}B, following the original paper. Specifically, its authors showed that each periodicity window is bordered by a saddle-node bifurcation curve SN$_{i-1,i}$ on one side such that a stable PO emerges upon inward-crossing of this curve, along with a saddle one. The stable orbit undergoes a cascade of period-doubling bifurcations. Curves PD$_{i-1,i}$ and PD$_{i,i+1}$ in Fig.~\ref{fig:Rossler_LZdiagr}B represent the first bifurcations in such cascades.

\begin{figure}[htb!]
\center{\includegraphics[width=1\linewidth]{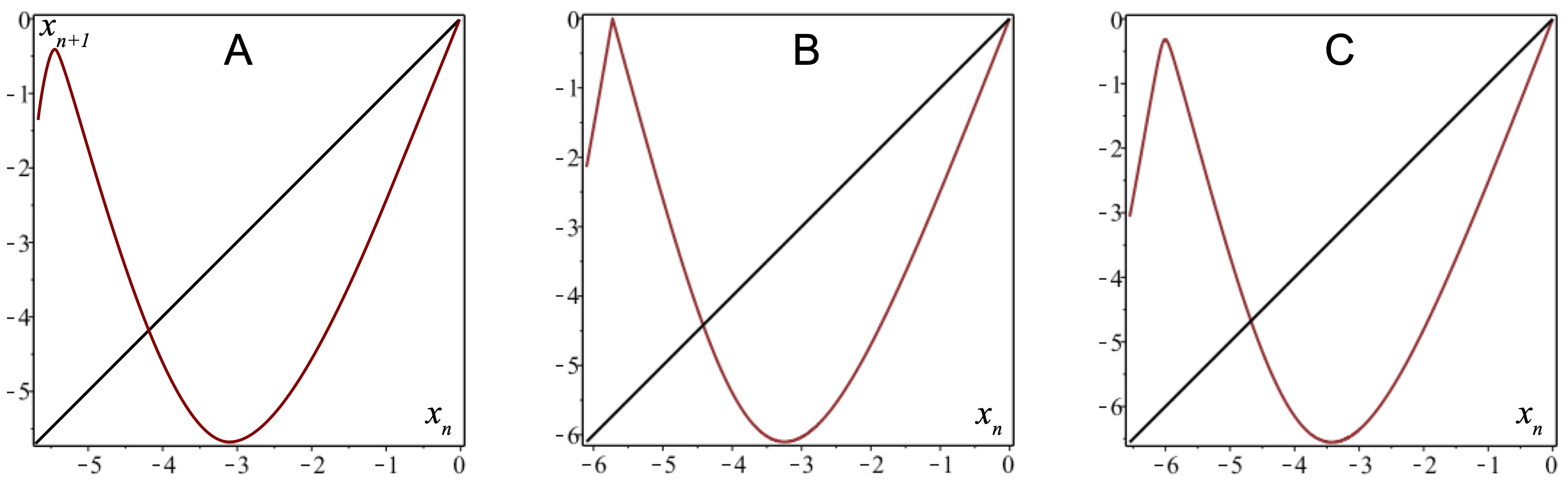}}
\caption{1D return map generated by the trajectories of the chaotic attractors sampled along the pathway $a=0.35$ near the primary bifurcation curve h$_1$: (A) the local minimum is below the repelling FP at the origin at $c = 4.6$ (on the left next to h$_1$), (B) the local minimum is taken to the origin after one iterate at  $c = 4.89$ (on h$_1$), and (C) the map at $c = 5.2$ (on the right next to h$_1$) is similar to the one in panel~A. }
\label{fig:1DMapsABC}
\end{figure}

\begin{figure*}[htb!]
\center{\includegraphics[width=0.8\linewidth]{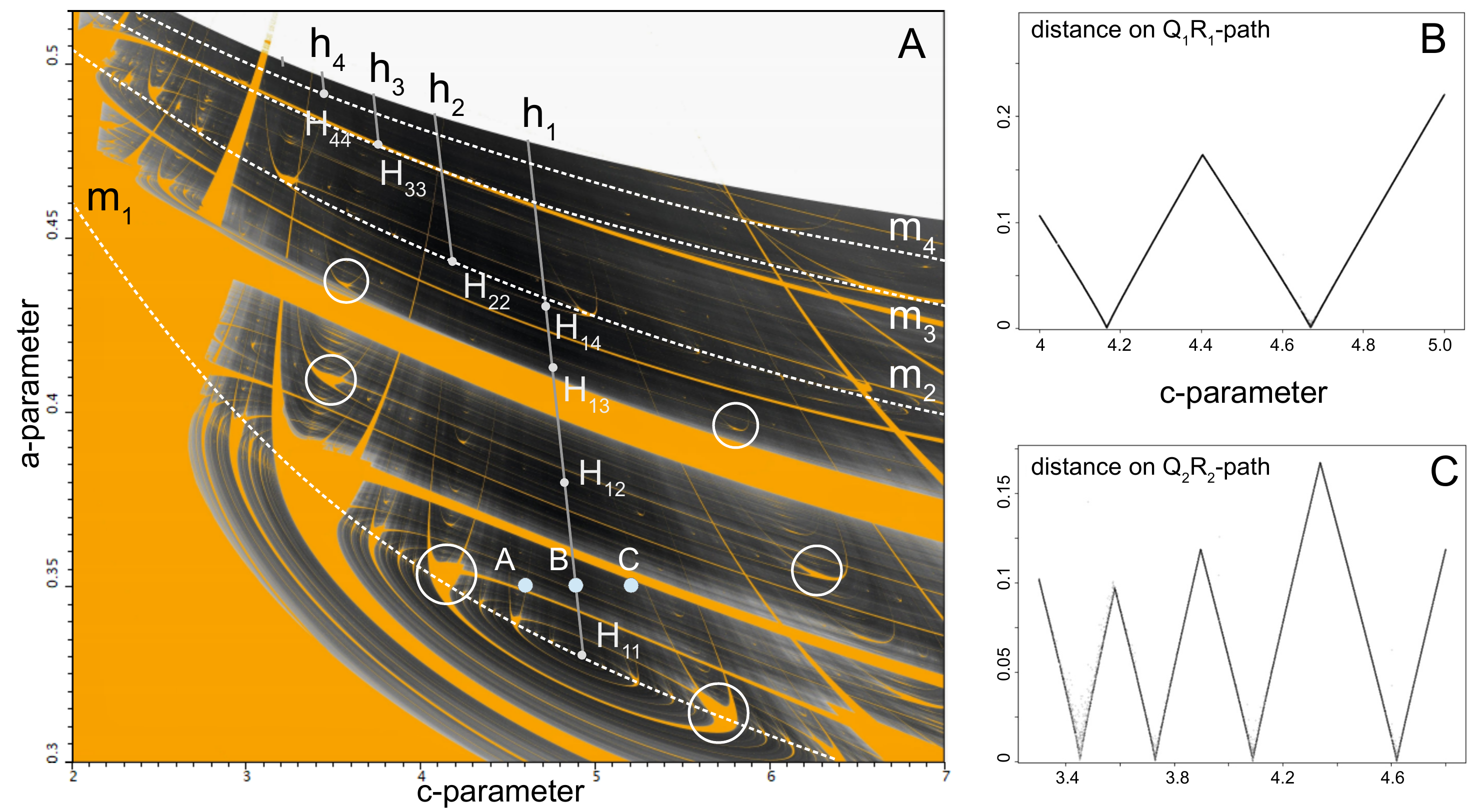}}
\caption{(A) Biparametric sweep with superimposed curves: m$_i$,  $(i=1,..,4)$ corresponding to the emerging $(i+1)$-th branches and new critical points in the 1D return maps, see examples in Figs.~\ref{fig:spiral_screw_funnel} and \ref{fig:primary_hubs}; specifically, above  m$_1$ the spiral attractor transforms into a screw-type~\cite{LetellierDutertreMaheu95, BarrioShilnikov2011}. Curves h$_i$ with tipping points H$_{ii}$ on m$_i$ stand for homoclinic bifurcations of $O_1$ and primary periodicity hubs H$_{ii}$, resp.  Shrimp-like windows (some circled) form what looks like nested spirals around the periodicity hubs. (B,\,C) the distance between the saddle-focus $O_1$ and the chaotic attractors on the P$_1$Q$_1$- and P$_2$Q$_2$-pathways, resp., plotted against the $c$-parameter. Its zeros correspond to the intersection points of these pathways with the indicated curves h$_i$. }
\label{fig:LZ_hubs}
\end{figure*}

Note a cusp point $c_i$ inside each shrimp-like stability window $S_i$, where the sub-criticality of SN-bifurcations changes due to such cusps, the dynamics inside  the shrimp-shaped region can be bistable. As was established in \cite{GaspardKapralNicolis84} that one curve, SN$_{12}$, originating from the cusp $c_1$ inside the shrimp S$_1$ connects to a subsequent cusp $c_2$ within the other shrimp S$_2$, containing the same stable POs. Furthermore, the bifurcation curve, SN$_{23}$, forming the left-bottom boundary of $S_2$ coalesces with the SN-curve representing the right-bottom boundary of the third shrimp $S_3$, and so forth. Due to such connectivity, these stability windows, alternating with regions of chaotic dynamics, appear as some nested spirals winding around what is code-named \textit{periodicity hubs}\cite{BonattoGallas08}, associated with a faux cod-2 bifurcation, if any. Note that this interesting structure was first reported in the original paper~\cite{GaspardKapralNicolis84}. Hubs in the R\"ossler model, as well as in other systems with homoclinic saddle-foci, were numerically studied using Lyapunov exponents in various diverse applications, for example see \cite{Gallas2010, VitoloGlendGallas2011, BonattoGallas08}, and the references therein.

It is interesting to note that, apparently, the SN and PD bifurcation curves edging the shrimp-shaped windows near the homoclinic bifurcation curve all branch out from a specific cod-2 point corresponding to a transition from a homoclinic saddle to a homoclinic saddle-focus.  In virtue of the Belyakov theorem \cite{Belyakov80}, such a point gives rise to a countable number of SN-curves and double homoclinic bifurcation curves. It is shown  in \cite{KuznetsovDeFeoRinaldi2001}, with the aid of the numerical parameter-continuation approach, that this is the case for the tritrophic food chain model with similar dynamical and bifurcational properties. As for the Rossler model, it is shown \cite{BarrioShilnikov2011} numerically that several SN and PD bifurcation curves do originate from the Belyakov point, as sketched in the top-chart in Fig.~\ref{fig:Rossler_LZdiagr}b.

Another interesting result reported in \cite{KuznetsovDeFeoRinaldi2001,BarrioShilnikov2011} is that all primary   and secondary homoclinic bifurcation curves in both food chain and R\"ossler models have the specific U-shaped form in the parameter space. Due to strong contraction and slow-fast dynamics, two close ($10^{-10}$) branches of the curves $h_1$ and others look like a single one, see a schematic representation  in Fig.~\ref{fig:Rossler_LZdiagr}C. The fold or turning point H$_{11}$ on h$_1$  represents the primary periodicity hub. Observe that that the left branch of h$_1$ corresponds to double homoclinic orbits, in contrast to orbits of the right branch \cite{KuznetsovDeFeoRinaldi2001,BarrioShilnikov2011}.

\subsection*{Primary homoclinic curves and periodicity hubs}

It was shown in \cite{VitoloGlendGallas2011} that on the left from the primary homoclinic bifurcation curve h$_1$, there exist several subsequent ones h$_i$, $i \in [2,3,\ldots]$ corresponding to double, triple and longer homoclinic orbits, see Figs.~\ref{fig:Rossler_BifDiagr}A and \ref{fig:LZ_hubs}A. All these curves are of the U-shaped form with fold points at the corresponding periodicity hubs H$_{ii}$. In Fig.~\ref{fig:LZ_hubs}, such curves can be typically detected and computed using numerical continuation toolkits such as AUTO and MatCont (in backwards time, making $O_1$ of (2,1)-type not (1,2)). We stress that in between curves h$_i$ and $h_{i+1}$, there are {\em no} other homoclinics to the saddle-focus~$O_1$. This fact can be simply verified using the proposed 1D maps. Let us consider three particular points in the $(c,a)$ bifurcation diagram: the points A~$(4.6,0.35)$ and C~$(5.2,0.35)$  are chosen on the opposite sides from  the bifurcation curve h$_1$ of the primary homoclinics of $O_1$, whereas the point B~$(4.89,0.35)$ is placed right on it, see Fig.~\ref{fig:LZ_hubs}A. At the point A, as well as the point C, the images of the local maximum in the 1D map are below the repelling FP at the origin $O$, which is interpreted as the saddle-focus $O_1$ is isolated, outside of the chaotic attractor, see Fig.~\ref{fig:1DMapsABC}A and \ref{fig:1DMapsABC}C. This is not the case for the point B, where the local maximum is taken to the FP $O$ after a single iterate (see \ref{fig:1DMapsABC}B), which indicates that the saddle-focus $O_1$ is the intrinsic component of this homoclinic attractor, in contrast to the other two cases.

In this work, we applied an alternative approach for finding the curves h$_i$ which is better suited for systems with various saddles of (1,2)-type, \cite{BKKLO18,KazKor18}. Our search strategy is based on the fact that a homoclinic attractor (the definition introduced in \cite{GGS12, GGKT14})) embeds homoclinic loops of $O_1$, whose simulated trajectories will come close by the saddle-focus $O_1$ eventually. By performing one- and two-parameter sweeps, we estimate the distance to $O_1$ from a long typical transient trajectory on the attractor. Whenever the distance becomes less than some threshold value $d_{tr}$ (we use $d_{tr}=0.001$), the homoclinic saddle-focus $O_1$ is meant to belong to the attractor.

Figure~\ref{fig:LZ_hubs}B is a de-facto demonstration of the efficiency of the described approach for two one-parameter pathways: P$_1$Q$_1$ given by $\{a = 0.45;\, 4\le c \le 5\}$) and P$_2$Q$_2$ given by $\{a = -1/60 c+0.55;\, 3.3 \le c \le 4.8\}$.  Along the first pathway P$_1$Q$_1$, the distance graph has two local minima below the threshold: the first minimum corresponds to the homoclinic bifurcation value on h$_2$ and the second one occurs on the curve h$_1$. Along the second pathway P$_2$Q$_2$, there are four such local minima in the graph, the distance plotted against the $c$-parameter, corresponding to the curves h$_4$, h$_3$, h$_2$, and h$_1$.

Next, let us describe the curves m$_1$, m$_2$, and m$_3$ that are also superimposed with the DCP-sweep in Fig.~\ref{fig:LZ_hubs}{\bf A}. These curves were found using the 1D return maps, namely: below m$_j$, 1D maps consist of $j$ connected parabolas, whereas above it, the maps gain the additional  ($j+1$th) increasing branch, as seen in Figs.~\ref{fig:spiral_screw_funnel}A$_2$ and \ref{fig:spiral_screw_funnel}B$_2$ depicting the 1D maps below and above the curve $m_1$, respectively \cite{KazKor18}. When the image of the new local maximum (on the far left branch) on the curve $m_j$  is the repelling FP $O$,  this corresponds to the occurrence of the $j$-round homoclinic orbit to the saddle-focus $O_1$. For example, see the 1D map at the hub H$_{11}$ on the primary curve h$_1$ through which the curve m$_1$ passes in Fig.~\ref{fig:mapScheme}.

\begin{figure}[t!]
\center{\includegraphics[width=1\linewidth]{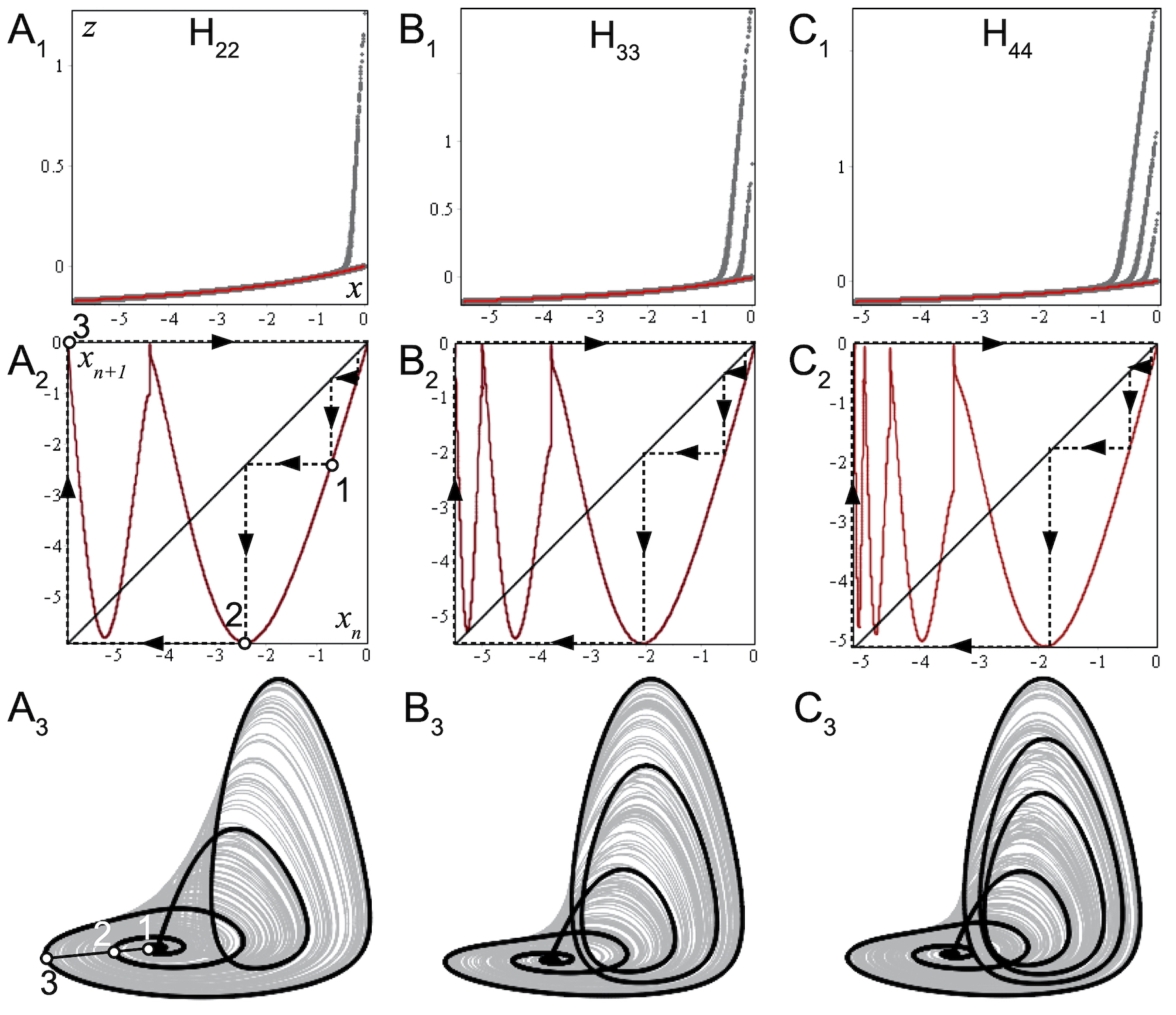}}
\caption{Homoclinic bifurcations at three primary hubs: (A) H$_{22}$ located at $(c \approx 4.182,  a \approx 0.4436)$, (B) H$_{33}$ at $(3.758,0.4777)$, and (C) H$_{44}$ around $(3.458,0.4916)$ (see the bifurcation diagram in Fig.~\ref{fig:LZ_hubs}A.) The corresponding 2D cross-sections (top panels), and 1D return maps (middle panels) for h$_i$-curves with H$_{ii}$-tuning points reveal $i$ branches  and parabolas, resp. with the saddle-focus $O_1$ placed in the top-right corner of the Lammerey (coweb) digram (A$_2$--C$_2$). Bottom panels illustrate the homoclinic orbits demarcating the ``edge'' of the chaotic attractor.}
\label{fig:primary_hubs}
\end{figure}

Figure~\ref{fig:primary_hubs} illustrates the homoclinic phenomena in other found hubs H$_{22}$, H$_{33}$, and H$_{44}$. The first row in this figure presents the 2D Poincar\'e maps computed at these points. Note that these maps contrast significantly from the map at  the primary hub H$_{11}$ (Fig.~\ref{fig:mapScheme}A), as they include 2, 3, and 4 branches, correspondingly, on the right-hand side.

Next, we parameterize these 2D maps by 1D ones.  To do that, we take four points on the bottom branch of each 2D map to compose the Lagrange polynomial, as previously described in Sec.~\ref{sec:1DMaps}. The resulting 1D  maps are shown in the middle row in Fig.~\ref{fig:primary_hubs}. One can see that the 1D map at H$_{ii}$-hub indeed consists of $i$ parabolas whose far left point is taken at the repeller $O$, after one iteration. The dotted lines representing  homoclinic obits in these Lammerey (coweb) diagrams correspond to the homoclinic orbits (bottom row) to the saddle-focus $O_1$ shown in the phase space of the R\"ossler model.

\begin{figure}[t!]
\center{\includegraphics[width=.6\linewidth]{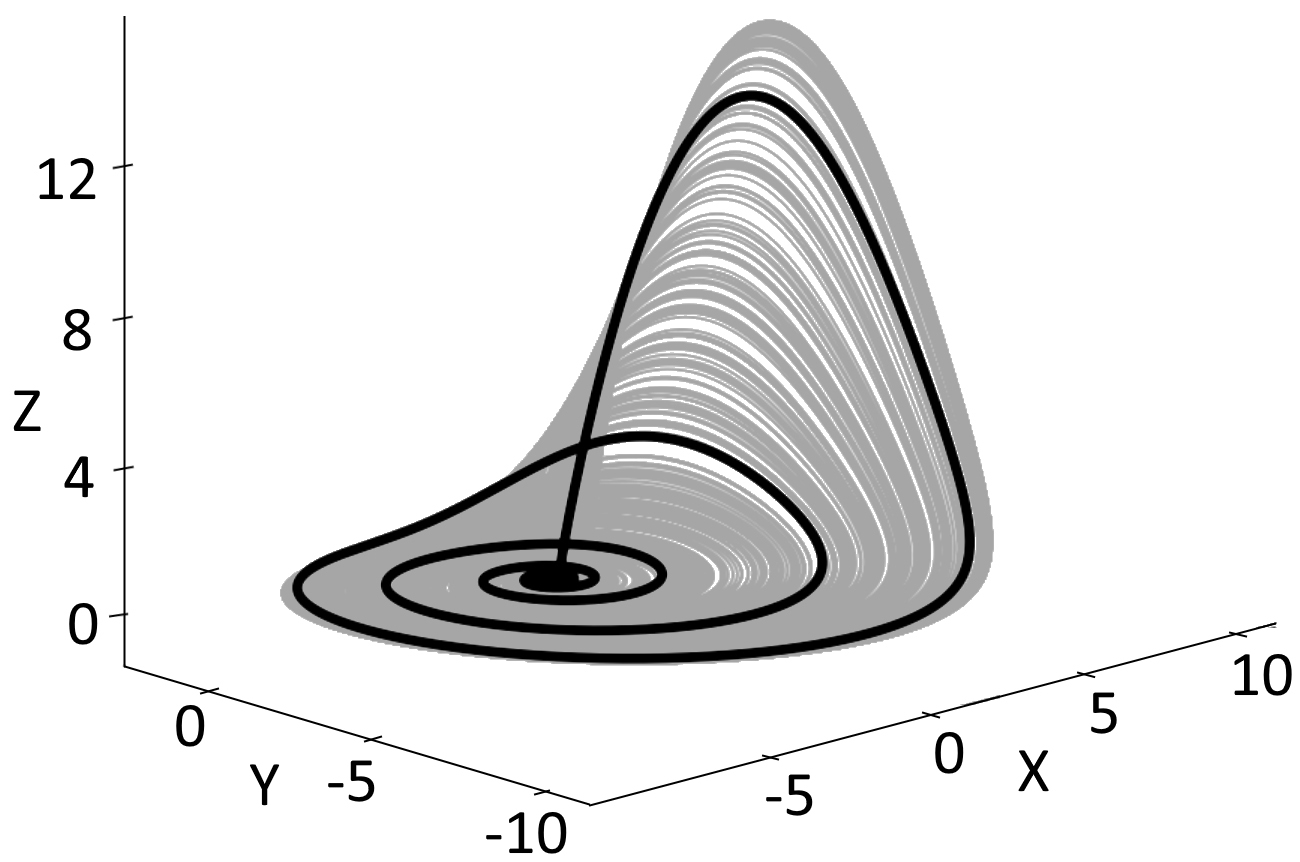}}
\caption{The separatrix loop of the saddle-focus $O_1$ at $(c=4.89,\, a=0.35)$ on the h$_1$-curve is not fully maximized to edge the homoclinic attractor (shown in the background), see the contrast with Figs.~\ref{fig:primary_hubs} and \ref{fig:secondary_hubs} for the  primary and secondary hubs, resp.}
\label{fig:AttrLoopB}
\end{figure}

It is important to underline another feature of periodicity hubs. In each hub the separatrix loop to  the saddle-focus is maximized in its size so that it becomes the ``edge'' of the homoclinic attractor, see Figs.~\ref{fig:mapScheme}B, \ref{fig:primary_hubs}A$_3$, \ref{fig:primary_hubs}B$_3$, and \ref{fig:primary_hubs}C$_3$. This is also documented in the corresponding 1D maps. Indeed, below H$_{ii}$ on $m_i$ the left branch of the 1D map remains lower than the FP $O$.  Above H$_{ii}$, a newly emerging branch is yet lower than the FP $O$, so the homoclinic orbit used it only to pass through before landing in the repelling FP at the origin. In contrast,
Fig.~\ref{fig:AttrLoopB} illustrates the homoclinic attractor with the superimposed homoclinic orbit  at the point B on the curve h$_1$, which is far from forming its edge.

\subsection{Hubs inside the U-shaped h$_1$-curve}

By carefully examining the bi-parameter sweeps, one can observe a hierarchy of secondary hubs (e.g. H$_{12}$, H$_{13}$, H$_{14}$) forming organization centers for shrimp-like stability windows, see Fig.~\ref{fig:LZ_hubs}A. As was shown in \cite{VitoloGlendGallas2011}, these secondary hubs lie inside the U-shaped curves h$_i$ corresponding to the primary homoclinic loops. Note
that the distinctive U-shape of the bifurcation curves is a common feature of diverse applications with a homoclinic saddle-focus, see for example ~\cite{MR1770499, GGNT97, XPSh2010, medrano2005basic} and the references therein.  In this subsection, we show how 1D maps help to predict the form of homoclinic loops originating from the most visible secondary hubs located inside the curve h$_1$.

\begin{figure}
\center{\includegraphics[width=1\linewidth]{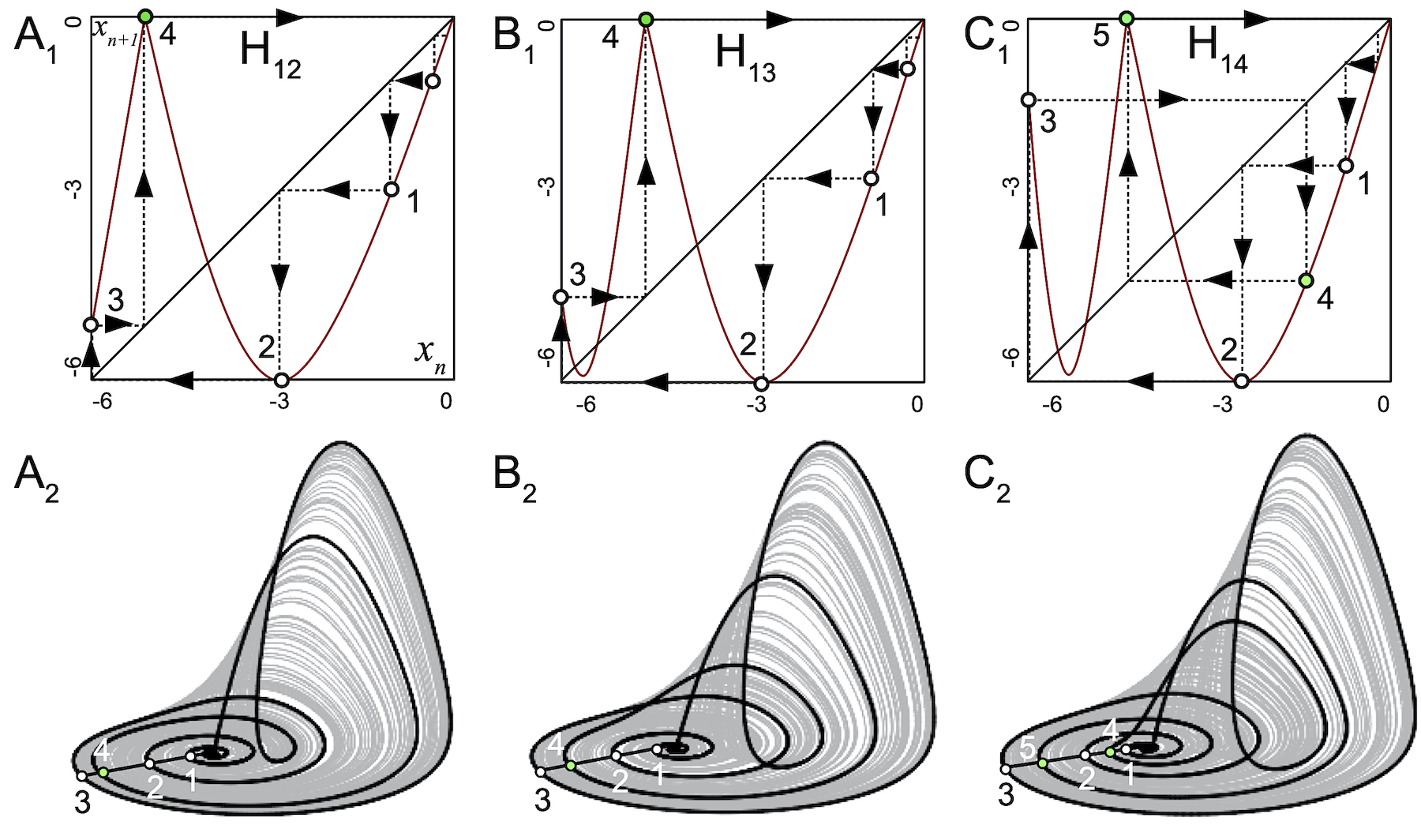} }
\caption{Homoclinic bifurcations at three secondary hubs: (A) H$_{12}$ located at $(c\approx 4.823,a \approx 0.38)$; (B) H$_{13}$ at $(4.753,0.4124)$; and H$_{14}$ near $(4.72,0.4279)$, see Fig.~\ref{fig:LZ_hubs}A. The corresponding simulated 1D maps depict the homoclinic orbits to the repeller FP $O$ located at the origin and associated with the saddle-focus $O_1$. Bottom panels depict the homoclinic loops edging the  superimposed chaotic attractors. }
\label{fig:secondary_hubs}
\end{figure}

Let us examine the fragment of h$_1$ between the curves m$_1$ and m$_2$ moving upward along h$_1$. While searching for the locations of secondary hubs in the diagram (Fig.~\ref{fig:LZ_hubs}A), let us analyze the corresponding 1D maps, while keeping in mind the important feature of all hubs, namely: in each hub the homoclinic orbit of the saddle-focus $O_1$ is {\em maximized} to ``edge'' the homoclinic attractor outwardly in the phase space, see  Figs~\ref{fig:primary_hubs} and \ref{fig:secondary_hubs} and contrast them with Fig.~\ref{fig:AttrLoopB}

The hubs H$_{12}$ is the most visible secondary hub along the curve h$_1$ (we mark the biggest shrimps around it by white circles in Fig.~\ref{fig:LZ_hubs}A). The corresponding 1D map is presented in Fig.~\ref{fig:secondary_hubs}A$_1$. The dotted lines represent homoclinic obits in the Lammerey (coweb) diagram. The corresponding homoclinic loop superimposed with the attractor is presented in Fig.~\ref{fig:secondary_hubs}A$_2$. Moving further along the curve h$_1$, the corresponding 1D map gains an additional decreasing branch. In the hub H$_{13}$, the far-left point of this branch matches with the repelling FP $O$ giving birth to a new homoclinic loop, see the corresponding 1D map with the Lammerey diagrams and phase portrait of the attractor with the homoclinic loop in  Figs.~\ref{fig:secondary_hubs}B$_1$ and ~\ref{fig:secondary_hubs}B$_2$, respectively.

In the H$_{14}$-hub below the curve m$_2$, the far-left decreasing branch gives rise to another homoclinic orbit passing on the edge of the attractor including, see the corresponding 1D map superimposed with the Lammerey diagram in Fig.~\ref{fig:secondary_hubs}C$_1$, and the phase projection of the corresponding homoclinic attractor presented in Fig.~\ref{fig:secondary_hubs}C$_2$.

Above the curve m$_2$, the 1D map gains an additional increasing branch and then, an additional decreasing one. Moving towards m$_3$, on the base of these branches, new homoclinic loops appear according to the described scenario. These loops make an additional global passage with respect to the loops in the hubs H$_{12}$, H$_{13}$, and H$_{14}$. Moreover, we believe that the same evolution of homoclinic loops is also observed along all curves h$_i$ corresponding to the primary i-round homoclinic loops.

Finally, we would like to emphasize that each periodicity hub H$_{ij}$ is a fold point for the corresponding homoclinic curve h$_{ij}$. Each such curve has a U-shape, with two extremely close to each other branches. All these curves reside inside the curve h$_i$ corresponding to the primary i-round homoclinic loop. The possible structure of the corresponding bi-parameter diagrams will be discussed in last section of this paper.

\begin{figure*}
\center{\includegraphics[width=.7\linewidth]{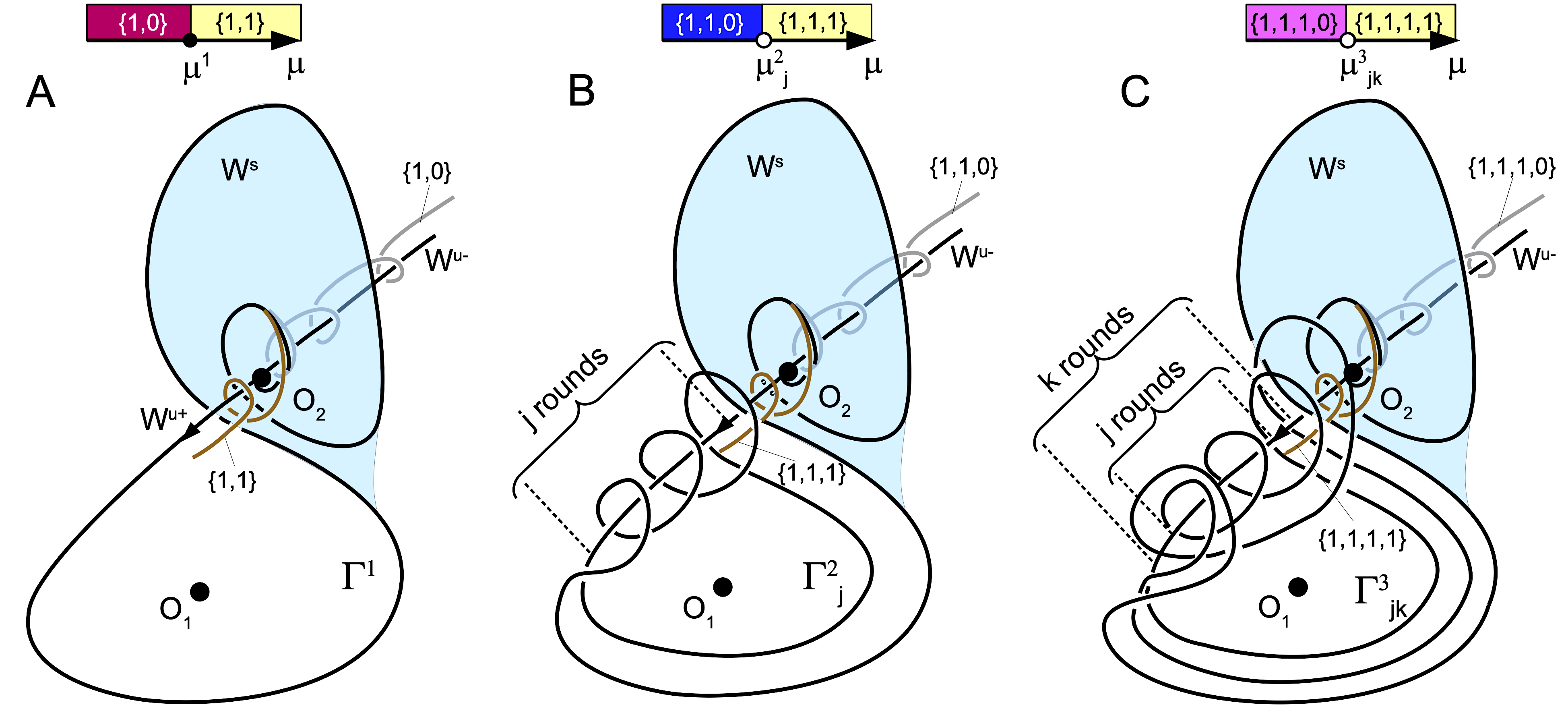} }
\caption{Geometric idea behind the symbolic algorithm for the detection of homoclinic loops to the saddle-focus $O_2$ with 2D stable manifold $W^s$ (shown in light blue) and 1D outgoing separatrices: returning $W^{u+}$ and $W^{u-}$ escaping to infinity. (A) Primary homoclinic orbit $\Gamma^1$ making a single turn around the saddle-focus $O_1$ at $\mu = \mu^0=0$ (splitting parameter)  is coded as $\{1\}$, or as $\{11\}$ if it makes it twice (B), or $\{111\}$ for three turns in (C), before it runs to infinity alongside $W^{u-}$ with a sequence $\{1110\}$. The notions $\Gamma^2_j$ and $\Gamma^3_{jk}$ for multiple double and triple homoclinic loops occurring at different values of $\mu^2_j$ and $\mu^2_{jk}$, resp., are used to indicate the number(s) $j,\,k$ of smaller rounds of $W^{u+}$ on the second and third turns before it returns to $O_2$. The further the system gets away from the primary loop, the longer the homoclinic orbits $\Gamma^N$ with increasing index $N$ become, with a greater multiplicity.}
\label{fig:KneadingScheme}
\end{figure*}

\section{Beyond the boundary of R\"ossler attractor} \label{sec:symbols}

This section reveals what the invisible role of the secondary saddle-focus $O_2$ is in the organization of the existence region for observable chaotic attractors in the R\"ossler model.  We will argue that the corresponding demarcation line above which the solutions start running to infinity (Fig.~\ref{fig:Rossler_LZdiagr}) can be well approximated by the curves of homoclinic bifurcations of $O_2$. The associated bifurcation curves are detected using the simulations combining the symbolic descriptions for homoclinic orbits of $O_2$ with biparametric sweeps.

It was shown in Sec.~\ref{sec:scenario} that on the demarcation line  the R\"ossler attractor merges with the 2D stable manifold  $W^s$ of the saddle-focus $O_2$ of (2,1)-type, see Fig.~\ref{fig:Attractor_StableManifold}. Recall that $O_2$ has two 1D unstable separatrices: $W^{u+}$ fills in the chaotic attractor, while  $W^{u-}$ always runs away to infinity. For the parameter values slightly above the demarcation line, the solutions start escaping along $W^{u-}$ (see Fig.~\ref{fig:Rossler_LZdiagr}H), because the 2D manifold $W^s$  of $O_2$ no longer shields them. This lets one say that the given crisis of the chaotic attractor is associated with an ``infinitely long'' homoclinic orbit of the saddle-focus $O_2$. As there is no way that such long homoclinics can be well computed, so we would limit ourselves to finding shorter homoclinic orbits and corresponding bifurcation curves in the $(c,\,a)$-diagram.  While sweeping $(c,a)$-plane, we calculate the number $N$ of global passages (or turns) of $W^{u+}$ around the primary saddle-focus $O_1$, before it crosses over $W^s$  of $O_2$ to escape to infinity.

\begin{figure*}[t]
\center{\includegraphics[width=.9\linewidth]{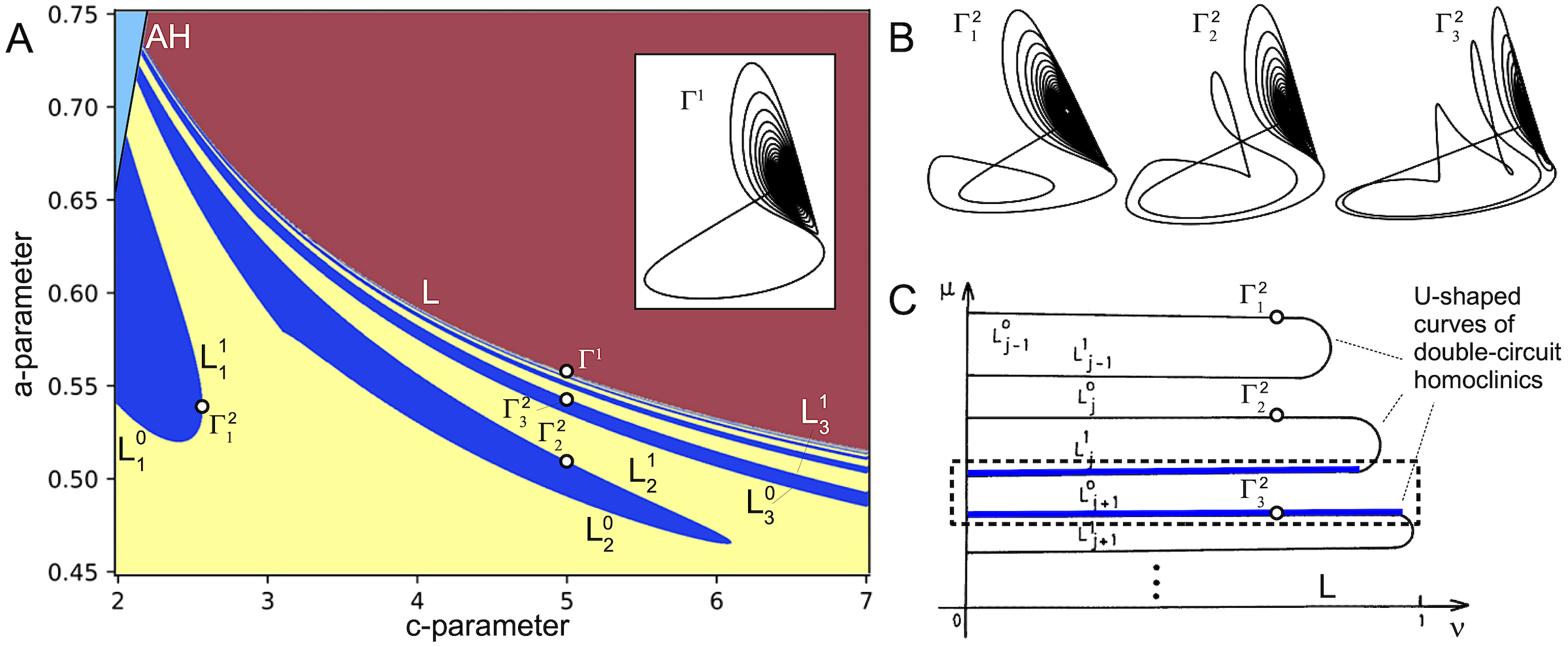} }
\caption{(A) ($c,a$)-parameter sweep capitalizing on the symbolic approach with binary sequences up to three symbols long: the red-colored regions is where the 1D unstable separatrix $W^{u+}$ of the saddle-focus $O_2$ runs to infinity after its first turn around $O_1$; the corresponding sequence is $\{1,0, 0\}$. The regions in blue/yellow colors correspond to the sequences $\{1,1,0\}$ and $\{1,1,1\}$, resp. The borderline $L$  between the red and yellow painted regions stands for the primary homoclinic loop $\Gamma^1$ of $O_2$, while boundaries (curves $L_j^0$ and $L_j^1$ merging in the some points) between the blue and yellow painted regions correspond to double homoclinic loops $\Gamma_j^2$, making $j$-turns  around $O_2$ on the second passage. (B) Homoclinic orbits $\Gamma_1^2$, $\Gamma_2^2$, and $\Gamma_3^2$ with one, two and three turns around the saddle-focus $O_2$ at specific points (white dots) in panel~A. (C) Bifurcation diagram, courtesy \cite{GGNT97}, in the $(\mu, \rho)$-parameter plane (here, $\mu$ measures the distance between a saddle-focus and its returning separatrix, and $\rho$ is the saddle-index introduced above). Horizontally stretching U-shaped curves $L_{j}^{0,1}$ for double homoclinic orbits with increasing $j$-index accumulate to the curve $L$ (given by $\mu=0$) corresponding to primary homoclinic orbit.}
\label{fig:2symbols}
\end{figure*}

\begin{figure*}
\center{\includegraphics[width=.9\linewidth]{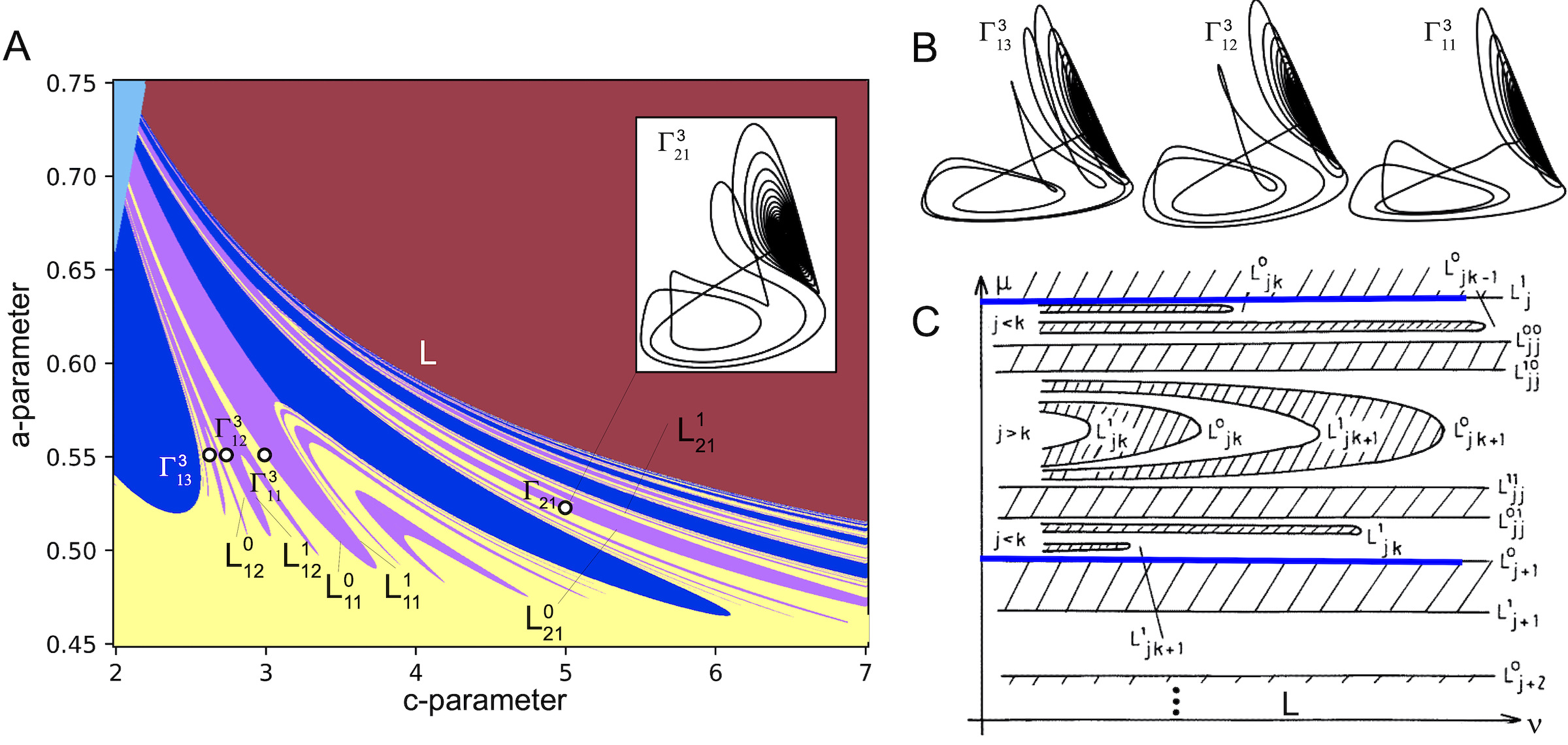} }
\caption{(A) Biparametric sweep revealing multiplicity of homoclinic orbits to the saddle-focus $O_2$ and the corresponding binary sequences up to $4$ symbols long. The color scheme is borrowed from Fig.~\ref{fig:2symbols}: the sequences $\{1,1,1,0\}$ correspond  are the borderlines of the regions  painted in the magenta color. The borderlines between the magenta and yellow colored regions (curves $L_{jk}^0$ and $L_{jk}^1$ bending at folds) correspond to the triple homoclinic loops $\Gamma_{jk}^3$ making $j$-rounds around $O_2$ on the second turn/passage and $k$-rounds on the third turn.  The boundary between the magenta and blue painted regions is not a bifurcation, as associated with transformations of double loops to the triple ones due to the integer arithmetics issue.  (B) Triple homoclinic loops $\Gamma_{13}^2$, $\Gamma_{12}^2$, and $\Gamma_{11}^2$ (with $1$ round on the first turn and $1$, $2$, and $3$ rounds on the third turn) in the three points marked in panel~A. (C) Schematic diagram, courtesy \cite{GGNT97}, revealing self-similar, nested organization of triple homoclinics curves located within a pair of curves $L_j^1$ and $L_{j+1}^0$, corresponding to double homoclinics.}
\label{fig:3symbols}
\end{figure*}

Specifically, for each parameter pair in the sweep, we generate a binary sequence $S$ according to the following algorithm. Whenever the unstable separatrix $W^{u+}$ of $O_2$ completes a global passage (or a sizable turn) around the equilibrium $O_1$, the symbol $1$ is added to $S$.  After that there are two options: (i) if $W^{u+}$ makes another turns around $O_1$, the second symbol $1$ is appended to the sequence $S=\{1,1\}$ and so forth $S=\{1,1,1,\dots\}$ ; (ii) if $W^{u+}$  runs to infinity, the second symbol is $0$ ($S=\{1,0\}$), and the calculations are halted for these parameter values. The latter case corresponds to the occurrence of the only primary  homoclinic orbit of the saddle-focus $O_1$, see Fig.~\ref{fig:KneadingScheme}A.  Unlike it, secondary, tertiary and longer homoclinics correspond to multiple bifurcation curves in the parameter plane.

Let us observe that, for example, between any $(c,\,a)$-parameter pair corresponding to sequences $\{1,1,1\}$ and $\{1,1,0\}$, there are always the pairs on some borderline, read a bifurcation curve,  corresponding to the double homoclinic orbits $\Gamma^2_{j}$, see Fig.~\ref{fig:KneadingScheme}B, wheres between sequences $\{1,1,1,1\}$ and $\{1,1,1,0\}$ there exists a triple homoclinic orbit $\Gamma^3_{jk}$, see Fig.~\ref{fig:KneadingScheme}C, and so on.

Finally, the pixels of the regions with unitary sequences $S=\{1,1,\ldots,1\}$ are painted in the yellow color, while regions corresponding  to different sequences ending with $0$, like ($S=\{1,1,\ldots,1,0\}$), are painted in contrast colors depending on the number of $1$'s in $S$, see the color bars at the top in Fig.~\ref{fig:KneadingScheme} for sequences of length $N = \{2, 3, 4\}$.

We begin our consideration with primary and double homoclinic orbits of the saddle-focus of $O_2$. Figure~\ref{fig:2symbols}A represents the corresponding biparametric sweep in the dedicated region $2 \le c \le 5 $ and $0.45 \le a \le 0.75$. The red-colored region is associated with the same sequence $S=\{1,0\}$, meaning that the unstable separatrix $W^{u+}$ runs to infinity after a single turn around the saddle-focus $O_1$. In the blue-colored regions associated with the same sequence $S=\{1,1,0\}$, $W^{u+}$  makes two turns around $O_1$ before escaping. In the yellow-painted region, it makes at least three turns around $O_1$ to generate  the sequence $S=\{1,1,1\}$.
Hence, the borderline between the red-colored region and the yellow-colored region corresponds to the primary homoclinic orbit $\Gamma^1$ of $O_2$.

The boundaries of the multiple blue-colored regions correspond to the double homoclinic orbits $\Gamma^2_{j}$, where the index $j$ indicates the number of small rounds around the outgoing separatrix  $W^{u+}$ of the saddle-focus $O_2$ on the second turn. As one can see from  Fig.~\ref{fig:2symbols}A, these bifurcation curves have a U-shaped form~\cite{Bel84}. We differentiate their branches with labels $L^0_j$ and $L^1_j$. It is known from \cite{Bel84, Gasp83} that these U-shaped curves of double loops accumulate to the primary curve $L$ \cite{GGNT97}, see Fig~\ref{fig:2symbols}C. The closer these curves approach to $L$, the greater the index $j$ becomes. One can see from the symbolic sweep presented in Fig.~\ref{fig:2symbols} that this agrees well with the theory. A few double homoclinic orbits $\Gamma^1_{j}$ are sampled in  Fig.~\ref{fig:2symbols}B for the indicated points on curves $L^1_{j}$, $j \in \{1,2,3\}$ in  Fig.~\ref{fig:2symbols}A.

Note that within the blue-colored regions, the unstable separatrix $W^{u+}$ always escapes after two turns around $O_1$. This implies that no other homoclinic bifurcations can populate these regions.  However, that is not the case for the yellow-colored region in which triple- and longer homoclinic orbits can occur. Fig.~\ref{fig:3symbols}A represents the homoclinic bifurcation diagram for the sequences of length up to four symbols. In it, there are multiple magenta-colored regions associated with the same sequence $S =\{1,1,1,0\}$ that populate the space between the blue-colored regions. Their boundaries, $L^0_{jk}$ and $L^1_{jk}$, correspond to triple homoclinic orbits $\Gamma^3_{jk}$, where the indices $j$ and $k$ stand, respectively, for the numbers of rounds on the second and third turns around $O_2$.

Bifurcations of triple homoclinic orbits $\Gamma^3_{jk}$ were studied in \cite{GGNT97}. The idea of bifurcation unfolding is sketched in Fig.~\ref{fig:3symbols}C. In the R\"ossler model, such bifurcations occur between curves $L^0_{j+1}$ and $L^1_{j}$ for double homoclinic orbits marked with blue horizontal lines in Fig.~\ref{fig:2symbols}C. Depending on the relationship between $j$ and $k$, these curves can be of either the U-shaped form if $j<k$ (as for double loops provided $j\leq k$), or of the form of horseshoe if $j > k$.

We would like to comment that specific boundaries between the magenta- and blue-colored regions in the symbolic sweep in Fig.~\ref{fig:3symbols}A are not bifurcation curves, but artifacts caused by  integer-based number of turns of $W^{u+}$ flying around $O_1$. Three triple homoclinic orbits $\Gamma^3_{jk}$ are sampled in Fig.~\ref{fig:3symbols}B for the marked parameter values on the curves $L^0_{jk}$, $j = 1$, $k \in \{1,2,3\}$ in the bifurcation diagram in Fig.~\ref{fig:3symbols}A.

\subsection*{Stitched bifurcation diagram}

\begin{figure}
\center{\includegraphics[width=1\linewidth]{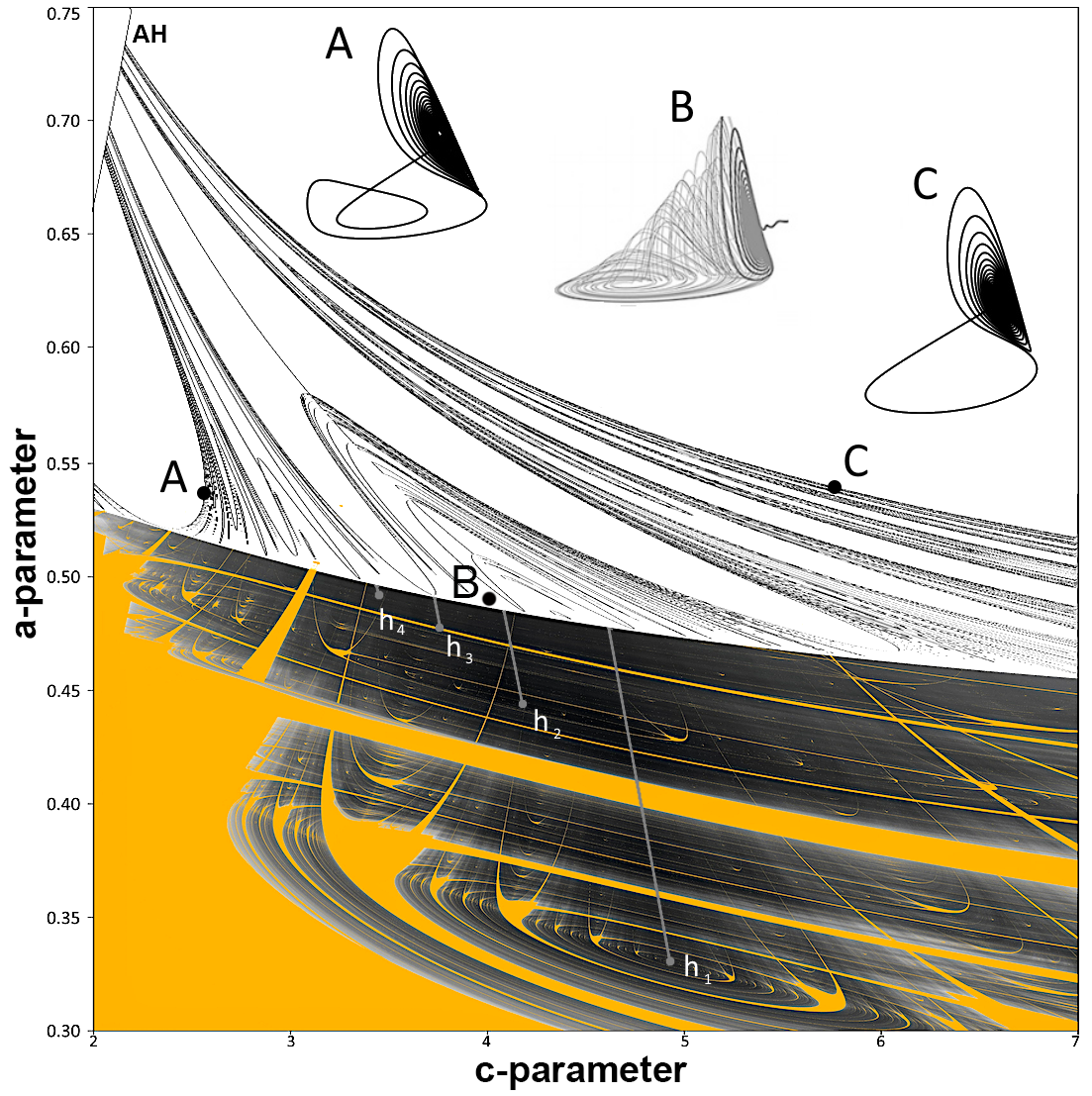}}
\caption{Stitched $(c,\,a)$-bifurcation diagram for the R\"ossler model. The top b/w chart is due to the primary (top line) and subsequent homoclinics  of the saddle-focus $O_2$ of (2,1)-type, whose corresponding bifurcation curves terminate on the sub-critical AH-bifurcation curve (that makes $O_2$ a repeller); the terminal points on the AH-curve represent the cod-2 Belyakov bifurcation nicknamed as Shilnikov-Hopf. The bottom part reflects the complex organization of regular and chaotic dynamics stirred up by the primary saddle-focus $O_1$ of (1,2)-type.}
\label{fig:JointSweep}
\end{figure}

Figure~\ref{fig:JointSweep} completing our case study of the R\"ossler model represents two bifurcation diagrams stitched together into one: the bottom section reveals the bifurcations mostly due to the primary saddle-focus $O_1$ in the phase space where the model is strongly dissipative, while the top chart includes the bifurcation curves corresponding to homoclinic orbits (up to the symbolic length 6) of the secondary, invisible saddle-focus $O_2$ in the phase subspace, with a positive divergence of the vector field. One can see from the stitched diagram that the U-shaped homoclinic curves well approximate from above, the demarcation line of the existence region of the R\"ossler attractors, regardless of whether they are regular or chaotic. For the sake of visual clarity, we did not use the longer homoclinic orbits in this chart.

We underline that without knowing the the homoclinic structures due to the secondary saddle-focus $O_2$, the complete perception of the origin and driving forces of the  observable chaotic dynamics in the R\"ossler model and the metamorphoses of its attractors stirred by the primary saddle-focus $O_1$, would be barely possible.

\section{$h_1$-interior hypothesis}\label{sec:h1Hyp}

In this last section, we would like to hypothesize about what way and order the h$_{1j}$-curves of the secondary homoclinics of the saddle-focus $O_1$ can be organized inside the primary bifurcation h$_1$-curve in the parameter space. The distance ($\sim 10^{-10}$) between the folded branches of h$_1$ is too infinitesimal to apply implicit computations at this scale, such as the parameter continuation.  Furthermore,  like any strong  dissipative system, solutions including  1D separatrices of $O_1$ without exception, of the R\"ossler model~\eqref{eq:Rossler},  while integrated in backward time, become highly sensitive to the smallest perturbations and tend to escape quickly to infinity in no time.

\begin{figure*}
\center{\includegraphics[width=.65\linewidth]{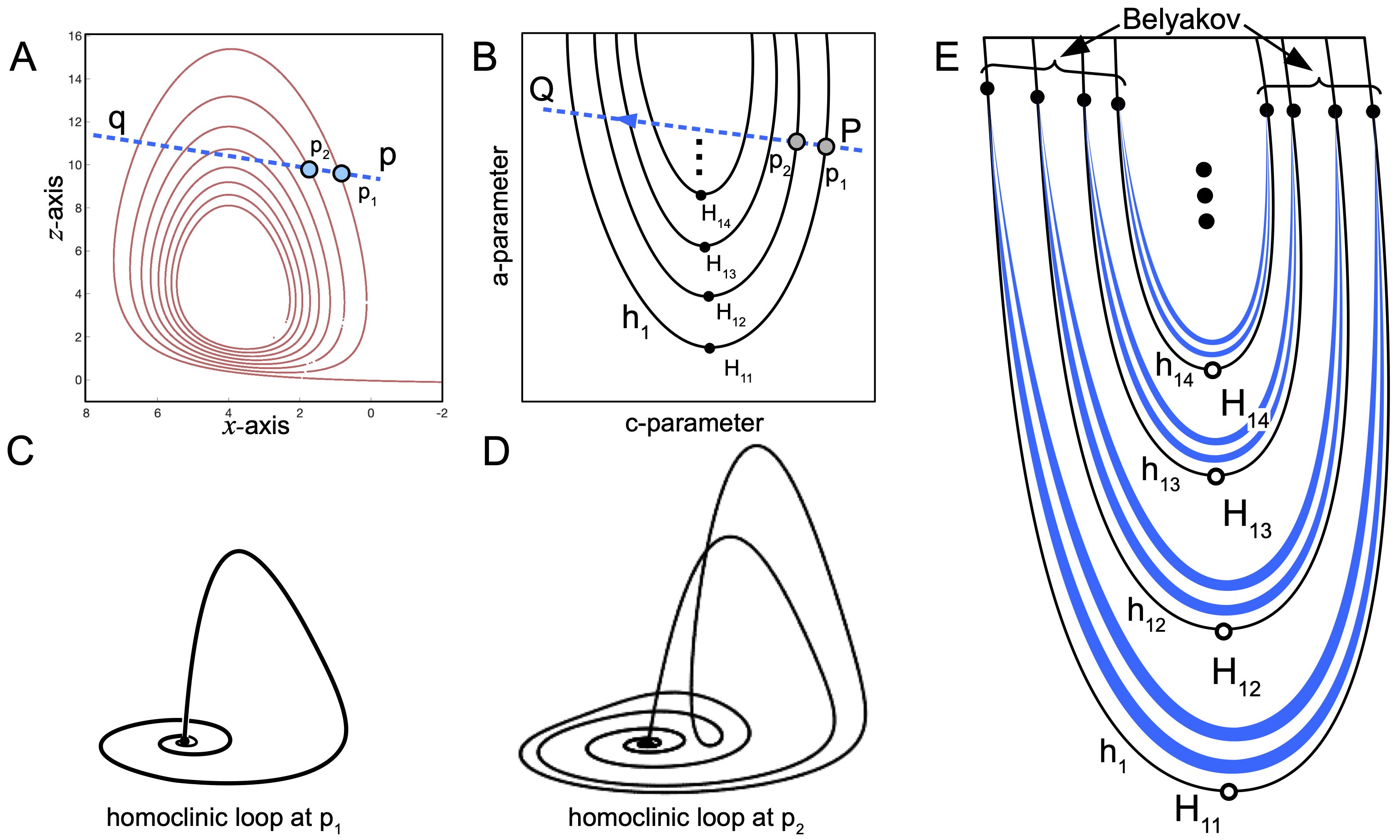}}
\caption{Inside the  h$_1$-curve.
 (A) Cross-cut through the multi-funnel attractor in Fig.~\ref{fig:Attractor_StableManifold}C reveales the spirals (orange) being the $W^u(O_1)$-image. Pathway labelled as $pq$ throughout $W^u$ is traced down by the first intersection point of $W^s(O_1)$ with the cross-section while following the local $PQ$-pathway transverse to h$_1$ in the $(c,\,a)$-diagram sketched in panel B. Two crossing points, p$_1$ and p$_2$, of the $W^s(O_1)$-image with the $W^u(O_1)$-image represent two homoclinic orbits shown in (C-D). (B) Sketch: U-shaped secondary h$_{12}$, h$_{13}$, and h$_{14}$-curves sequentially nested within each other and the primary h$_1$-curve; see the corresponding homoclinic orbits in Fig.~\ref{fig:secondary_hubs}. (E) The diagram connecting the Belyakov points with h$_{ij}$-curves from panel D.}
\label{fig:O1_scetch}
\end{figure*}

So, let us try to re-visit and examine further, the other not fully exploited options.  Fig.~\ref{fig:Attractor_StableManifold} is one such example. Recall that the orange spirals in it represent the image of the homoclinic attractor transversely cut through by a 2D cross-section.  In other words, by construction, spirals flattened in Fig.~\ref{fig:O1_scetch}A represent the computational point-wise reconstruction of the 2D unstable manifold $W^u$ of the saddle-focus $O_1$ in the same cross-section on which some 2D return map is thereby empirically defined. Note that the intersection(s) of the 1D $W^s$ of $O_1$ is some point(s), which are close to the spirals representing the $W^u(O_1)$-image the h$_1$-curve in the parameter space. The primary homoclinic orbit of $O_1$ emerges when $W^s(O_1)$ touches $W^u(O_1)$, or in the 2D map reduction, when the $W^s(O_1)$-image -- the first intersection point, namely the $p_1$-point in Fig.~ ~\ref{fig:O1_scetch}A,   touches the spiraling  $W^u(O_1)$-image.

 Let us make a virtual experiment: locally cross the h$_1$-curve twice by following the $PQ$-pathway from the right to the left, see Fig.~\ref{fig:O1_scetch}B.  By construction, crossing h$_1$ only or several h$_{1j}$-curves (within h$_1$) along the $PQ$-pathway in the $(c,a)$-plane makes the $W^s(O_1)$-image trace down a $pq$-pathway in the 2D map as well that crosses twice (in and out) one or consecutively several nested arches of the spiraling $W^u(O_1)$-image. Therfore, the first crossing at the point, p$_1$, in the map (Fig.~\ref{fig:O1_scetch}A) occurs when the h$_1$-curve is inward-intersected from right to left. The corresponding homoclinic orbit is presented in Fig.~\ref{fig:O1_scetch}C. The second intersection point, $p_2$ corresponds to the secondary loop occurring on the h$_{12}$-curve, see Fig.~\ref{fig:O1_scetch}D, and so forth.

 A similar reasoning is also applicable to disclose the structure of the periodicity hubs: whenever the parametric pathway passes right through a hub, then the $W^s(O_1)$-trace (read the $pq$-pathway) becomes only tangent to some spiral of the $W^u(O_1)$-image in the corresponding 2D map.
Now we can argue to justify a nested organization of the h$_{1j}$-curves.  The assertion is that the $W^s(O_1)$-trace typically (not in hubs) crosses each spiral or arch of the $W^u(O_1)$-image twice. Each pair of such crossing points corresponds to the two branches of each  U-shaped h$_{1j}$-curve nested successfully within a similar outer U-shaped  h$_{1j-1}$-curve, and so forth, as sketched in Fig.~\ref{fig:O1_scetch}B.
Panel~E in Fig.~\ref{fig:O1_scetch} presents our vision how a global reconstruction of h$_{1j}$-curves linking the Belyakov cod-2 points and the periodicity hubs in the parameter plane may possibly look like. Let us make it clear again that this is so far our best interpretation of this homoclinic puzzle based upon performed simulations and bifurcation logic. It would be helpful to compare our hypothesis with theoretical and computational findings from other systems featuring both periodicity hubs and Belyakov cod-2 bifurcations of the given type.

\section*{Conclusion}

In this paper, we presented a case study of essential homoclinic bifurcations two Shilnikov saddle-foci that govern the shape and the intrinsic structure of developed deterministic chaos observable in the R\"ossler model. We combine two computational approaches -- the first one capitalizing on 1D Poincar\'e return maps and the second utilizing the symbolic description to provide  both evident qualification and  proper quantification of underlying global bifurcations of the R\"ossler strange and periodic attractors. The map-based approach lets us accurately detect the location of the homoclinic bifurcation curves and their turning points -- the periodicity hubs in the parameter space. The symbolic approach with long transients is the new development in the computational apparatus in nonlinear sciences that aids to identify the regions of chaotic and periodic dynamics in biparametric sweeps. The use of short binary sequences is a powerful instrument to locate and investigate homoclinic saddle-focus bifurcation in the given model. The generality of our computational toolkit makes it universal and applicable to other systems of diverse origins, ranging from mathematics through life sciences.

\section*{Acknowledgements}
We thank the Brains and Behavior initiative of Georgia State University for the B\&B fellowship awarded to K. Pusuluri. We are grateful to  D. Turaev and J.~Scully for inspiring discussions, as well as H.G.E.  Meijer for  MatCont tutoring. The Shilnikov NeurDS lab thanks the NVIDIA Corporation for donating the Tesla K40 GPUs that were actively used in this study. A.~Kazakov and A.~Shilnikov  acknowledge a partial funding support from the Laboratory of Dynamical Systems and Applications NRU HSE, grant No. 075-15-2019-1931 from the Ministry of Science and Higher Education of Russian Federation. Yu.~Bakhanova and A.~Kazakov acknowledge the RSF grant No. 19-71-10048 for the funding support related to the results presented in Sec.~IV and Sec.~VI. S.~Malykh acknowledges the RSF grant No. 20-71-10048 for the funding support related to the results presented in Sec.~V.

\section*{Data/Code Availability}
The DCP code is open source and freely available at \url{https://bitbucket.org/pusuluri_krishna/deterministicchaosprospector/}.

\section*{References}


\providecommand{\noopsort}[1]{}\providecommand{\singleletter}[1]{#1}%

\end{document}